\documentclass[runningheads,hidelinks,doubleblind]{llncs}
\usepackage{subcaption} 
                        
\usepackage{amsmath,amsfonts, amssymb}

\usepackage{graphicx}
\usepackage{nameref}
\usepackage{stmaryrd}
\usepackage[usenames,dvipsnames]{xcolor}
\usepackage{xfrac}
\usepackage{xspace}
\usepackage[shortlabels]{enumitem}
\usepackage{cancel}
\usepackage{csquotes}
\usepackage{adjustbox}
\usepackage{nicefrac}
\usepackage{mathtools}
\usepackage{booktabs}
\usepackage{multirow}
\usepackage{listings}
\usepackage{marvosym}
\usepackage{fontawesome}
\usepackage[ruled,vlined,linesnumbered]{algorithm2e}
\SetKwRepeat{Do}{do}{while}

\SetCommentSty{mycommfont}

\usepackage{tensor}
\usepackage{tikz}
\usetikzlibrary{patterns,arrows,backgrounds,calc,shapes,shadows,decorations.pathmorphing,decorations.pathreplacing,automata,shapes.multipart,positioning,shapes.geometric,fit,circuits,trees,shapes.gates.logic.US,fit,decorations.markings}

\usepackage[font=small,labelfont=bf]{caption}

\newtheorem{counterexample}{Counterexample}

\makeatletter
\def\my@tag@font{\small}
\def\maketag@@@#1{\hbox{\m@th\normalfont\my@tag@font#1}}
\let\amsmath@eqref\eqref
\renewcommand\eqref[1]{{\let\my@tag@font\relax\amsmath@eqref{#1}}}
\makeatother

\usepackage{scalerel} 

\newcommand{\tool}{\textsc{kipro2}}
\newcommand{\expname}[1]{\textnormal{\texttt{#1}}}

\renewcommand{\inf}{\ensuremath{\!\bigsqcap}}

\newcommand{\pstate}{\ensuremath{\sigma}\xspace}

\newcommand{\bb}{\ensuremath{\varphi}\xspace}
\newcommand{\bp}{\ensuremath{\psi}\xspace}
\newcommand{\br}{\ensuremath{\rho}\xspace}

\newcommand{\rr}{\ensuremath{r}\xspace}

\makeatletter
\newcommand{\monus}{\mathbin{\text{\@dotminus}}}
\newcommand{\@dotminus}{%
	\ooalign{\hidewidth\raise1ex\hbox{.}\hidewidth\cr$\m@th-$\cr}%
}
\makeatother

\newcommand{\qflira}{\textsf{QF\kern-.7pt\_LIRA}\xspace}
\newcommand{\qfuflira}{\textsf{QF\kern-.7pt\_UFLIRA}\xspace}

\newcommand{\smtentails}[2]{\ensuremath{\textsf{cex}_{\preceq}\left( #1 , #2 \right)}\xspace}
\newcommand{\tosmt}[1]{\ensuremath{\textsf{encodeInfty} \left( #1\right)\xspace}}
\newcommand{\smtinfty}{\ensuremath{\textsf{infty}}\xspace}

\newcommand{\sfsymbol}[1]{\textsf{\upshape {#1}}}

\newcommand{\ttsymbol}[1]{\texttt{\upshape {#1}}}

\newcommand{\wpsymbol}{\sfsymbol{wp}}

\renewcommand{\wp}[2]{\wpsymbol\llbracket #1\rrbracket\left(#2\right)}

\newcommand{\conditionalPair}[2]{{\let\oldarraystretch\arraystretch}\renewcommand{\arraystretch}{1}~\holter{~\raisebox{.5ex}{${#1}$}~}{~\raisebox{.125ex}{${#2}$}~}~\renewcommand{\arraystretch}{\oldarraystretch}}

\newcommand{\cc}{\ensuremath{C}} 
\newcommand{\cloop}{\cc_{\textnormal{loop}}} 
\newcommand{\cgeo}{\cc_{\textnormal{geo}}} 
\newcommand{\cbody}{\cc} 
\newcommand{\guard}{\bb} 
\newcommand{\ee}{\ensuremath{e}} 
\newcommand{\eeinfty}{\ensuremath{\tilde{e}}} 
\newcommand{\eainfty}{\ensuremath{\tilde{a}}} 
\newcommand{\eoinfty}{\ensuremath{\tilde{d}}} 

\newcommand{\pp}{\ensuremath{p}} 

\newcommand{\ff}{\ensuremath{f}} 
\newcommand{\fg}{\ensuremath{g}}
\newcommand{\fh}{\ensuremath{h}}
\newcommand{\fhp}{\ensuremath{h'}}

\newcommand{\oa}{\ensuremath{\alpha}} 
\newcommand{\ob}{\ensuremath{\beta}} 
\newcommand{\oc}{\ensuremath{\gamma}} 
\newcommand{\od}{\ensuremath{\delta}} 

\newcommand{\SKIP}{\ttsymbol{skip}}

\newcommand{\AssignSymbol}{\mathrel{\textnormal{$\mathtt{\coloneqq}$}}}
\newcommand{\ASSIGN}[2]{\ensuremath{#1 \AssignSymbol #2}}

\newcommand{\AVAILLOC}[1]{\PosNats}
\usepackage{actuarialangle}

\newcommand{\COMPOSE}[2]{\ensuremath{{#1}{\,;}~ {#2}}}
\newcommand{\PCHOICE}[3]{\ensuremath{\left\{\, {#1} \,\right\}\mathrel{\left[\,#2\,\right]}\left\{\, {#3} \,\right\}}}

\newcommand{\IFSYMBOL}{\ensuremath{\textnormal{\texttt{if}}}}

\newcommand{\ELSESYMBOL}{\ensuremath{\textnormal{\texttt{else}}}}

\newcommand{\ITE}[3]{\ensuremath{\IFSYMBOL\,\left(\, {#1} \,\right)\,\left\{\, {#2} \,\right\}\,\ELSESYMBOL\,\left\{\, {#3} \,\right\}}}

\newcommand{\WHILESYMBOL}{\ensuremath{\textnormal{\texttt{while}}}}
\newcommand{\WHILE}[1]{\ensuremath{\WHILESYMBOL \left(\, {#1} \,\right)\left\{\right.}}

\newcommand{\WHILEDO}[2]{\ensuremath{\WHILESYMBOL \left(\, {#1} \,\right)\left\{\, {#2} \,\right\}}}

\newcommand{\TICK}[1]{\ensuremath{\textnormal{\texttt{tick}} \left( #1 \right)}}

\newcommand{\varfont}[1]{\textit{#1}}
\newcommand{\varsent}{\ensuremath{\varfont{sent}}\xspace}
\newcommand{\varfailed}{\ensuremath{\varfont{fail}}\xspace}
\newcommand{\vartosend}{\ensuremath{\varfont{toSend}}\xspace}
\newcommand{\varmaxfailed}{\ensuremath{\varfont{maxFail}}\xspace}
\newcommand{\vartotalfailed}{\ensuremath{\varfont{totalFail}}\xspace}

\newcommand{\statepred}[1]{\ensuremath{\langle #1 \rangle}\xspace}

\newcommand{\refute}{\textsf{refute}}
\newcommand{\verify}{\textsf{verify}}

\newcommand{\sqsupsetneq}{\sqsupset}
\newcommand{\sqsubsetneq}{\sqsubset}

\newcommand{\laset}{\ensuremath{E}}
\newcommand{\laord}{\ensuremath{\sqsubseteq}}
\newcommand{\lambient}{\ensuremath{\nu}}
\newcommand{\llaord}{\ensuremath{~{}\sqsubseteq{}~}}

\newcommand{\llaordrev}{\ensuremath{~{}\sqsupseteq{}~}}

\newcommand{\lameet}{\sqcap}

\newcommand{\laf}{\ensuremath{f}}
\newcommand{\lag}{\ensuremath{g}}
\newcommand{\lah}{\ensuremath{h}}

\newcommand{\lak}{\ensuremath{\kappa}}

\newcommand{\laop}{\ensuremath{\Phi}}

\newcommand{\laup}[1]{\ensuremath{\laop^{\left\lceil #1 \right\rceil}_{\vphantom{g}}}}
\newcommand{\ladown}[1]{\ensuremath{\laop^{\left\lfloor #1 \right\rfloor}}}

\newcommand{\lapp}[1]{\ensuremath{\laop\left( #1 \right)}}
\newcommand{\lappup}[2]{\ensuremath{\laup{#1}\left( #2 \right)}}
\newcommand{\lappdown}[2]{\ensuremath{\ladown{#1}\left(#2\right)}}
\newcommand{\auxdown}[2]{\ensuremath{\Lambda^{\lfloor #1 \rfloor}\left(#2\right)}}

\newcommand{\kindop}[1]{\ensuremath{\Psi_{#1}}}
\newcommand{\kindopfold}[2]{\ensuremath{\Psi_{#1}^{\left\lfloor #2 \right\rfloor}}}
\newcommand{\kindopapp}[2]{\ensuremath{\Psi_{#1}\left( #2 \right)}}

\newcommand{\pgcl}{\textnormal{\sfsymbol{pGCL}}\xspace}   
\newcommand{\Vars}{\ensuremath{\mathsf{Vars}}\xspace}   

\newcommand{\init}{\ensuremath{I}\xspace}
\newcommand{\initf}{\ensuremath{I}\xspace}
\newcommand{\transsys}{\textnormal{TS}\xspace}
\newcommand{\transrel}{\ensuremath{T}\xspace}
\newcommand{\transrelf}{\ensuremath{T}\xspace}
\newcommand{\stateset}{\ensuremath{S}\xspace}
\newcommand{\propset}{\ensuremath{P}\xspace}
\newcommand{\propsetf}{\ensuremath{P}\xspace}
\newcommand{\succs}{\textnormal{\textsf{Succs}}\xspace}
\newcommand{\succsk}[2]{\ensuremath{\succs^{\lfloor #1 \rfloor}\left(#2\right)}}

\newcommand{\Nats}{\ensuremath{\mathbb{N}}\xspace}

\newcommand{\PosNats}{\ensuremath{\mathbb{N}_{>0}}\xspace}
\newcommand{\Ints}{\ensuremath{\mathbb{Z}}\xspace}
\newcommand{\Rats}{\ensuremath{\mathbb{Q}}\xspace}

\newcommand{\PosReals}{\mathbb{R}_{\geq 0}}
\newcommand{\PosRats}{\mathbb{Q}_{\geq 0}}
\newcommand{\PosRealsInf}{\mathbb{R}_{\geq 0}^\infty}

\newcommand{\onetoarg}[1]{\ensuremath{\mathbf{#1}}\xspace}

\newcommand{\E}{\mathbb{E}}
\newcommand{\eleq}{\preceq}
\newcommand{\SyntE}{\ensuremath{\mathsf{Exp}}\xspace}
\newcommand{\Elin}{\ensuremath{\mathsf{LinExp}}\xspace}

\newcommand{\gnf}[1]{\ensuremath{\mathsf{GNF} \left( #1\right)}\xspace}

\newcommand{\emin}{\ensuremath{\mathrel{\mathsf{min}}} \xspace}

\newcommand{\iverson}[1]{\left[ {#1} \right]}

\newcommand{\subst}[2]{\left[ {#1} \middle/ {#2}\right]}
\newcommand{\statesubst}[2]{\left[ {#1} \mapsto {#2}\right]}

\newcommand{\States}{\Sigma}

\newcommand{\true}{\mathsf{true}}
\newcommand{\false}{\mathsf{false}}
\newcommand{\mydot}{\text{{\Large\textbf{.}}~}}

\newcommand{\tiff}{\textnormal{iff}\quad}
\newcommand{\qiff}{~\quad\textnormal{iff}\quad~}
\newcommand{\qqiff}{\qquad\textnormal{iff}\qquad}

\newcommand{\qand}{~\quad\textnormal{and}\quad~}

\newcommand{\qimplies}{~\quad\textnormal{implies}\quad~}
\newcommand{\qqimplies}{\qquad\textnormal{implies}\qquad}

\newcommand{\ppreceq}{~{}\preceq{}~}

\newcommand{\defeq}{{}\triangleq{}}
\newcommand{\ddefeq}{~{}\defeq{}~}
\newcommand{\eeq}{~{}={}~}
\newcommand{\mmapsto}{~{}\mapsto{}~}

\newcommand{\qmid}{\quad{}|{}\quad}
\newcommand{\tto}{~{}\to{}~}

\newcommand{\qeq}{\quad{}={}\quad}

\newcommand{\lleq}{~{}\leq{}~}
\newcommand{\LL}{~{}<{}~}

\newcommand{\wwedge}{~{}\wedge{}~}

\newcommand{\set}[1]{\left\{\, {#1} \,\right\}}
\newcommand{\setcomp}[2]{\left\{\, {#1} ~\middle|~ {#2} \,\right\}}

\newcommand{\underdot}[1]{%
    \tikz[baseline=(todotted.base)]{
        \node[inner sep=1pt,outer sep=0pt] (todotted) {#1};
        \draw[densely dotted] (todotted.south west) -- (todotted.south east);
    }%
}%
\newcommand{\cloze}[1]{\underdot{\phantom{#1}}}

\newcounter{computationarrowsone}
\newcounter{computationarrowstwo}

\def\triangleforqed{\hbox{$\lhd$}}
\makeatletter
\DeclareRobustCommand{\exped}{%
	\ifmmode
	\eqno \def\@badmath{$$}
	\let\eqno\relax \let\leqno\relax \let\veqno\relax
	\hbox{\triangleforqed}%
	\else
	\leavevmode\unskip\penalty9999 \hbox{}\nobreak\hfill
	\quad\hbox{\triangleforqed}%
	\fi
}
\makeatother

\newcounter{sarrow}

\newcommand{\lfp}{\ensuremath{\textnormal{\sfsymbol{lfp}}~}}
\newcommand{\gfp}{\ensuremath{\textnormal{\sfsymbol{gfp}}~}}

\usepackage[textwidth=4cm]{todonotes}

\makeatletter
\def\thanks#1{\protected@xdef\@thanks{\@thanks
		\protect\footnotetext{#1}}}
\makeatother

\makeatletter
\RequirePackage[bookmarks,unicode,colorlinks=true]{hyperref}%
\def\@citecolor{blue}%
\def\@urlcolor{blue}%
\def\@linkcolor{RedViolet}%

\def\orcidID#1{\smash{\href{http://orcid.org/#1}{\protect\raisebox{-1.25pt}{\protect\includegraphics{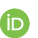}}}}}
\makeatother

\usepackage{cleveref}
\Crefname{requirement}{Requirement}{Requirements}

\allowdisplaybreaks

\begin{document}

\addtocontents{toc}{\protect\setcounter{tocdepth}{-1}}

\title{Latticed $k$-Induction \\ with an Application to Probabilistic Programs\thanks{\setlength{\leftskip}{0em}%
		This work has been partially funded by the ERC Advanced Project FRAPPANT under grant No.~787914.
	}
}



\authorrunning{K.~Batz et al.}

\author{Kevin Batz\inst{1}$^{\text{(\Letter)}}$\orcidID{0000-0001-8705-2564}
\and
Mingshuai Chen\inst{1}$^{\text{(\Letter)}}$\orcidID{0000-0001-9663-7441}
\and
Benjamin Lucien Kaminski\inst{2}$^{\text{(\Letter)}}$\orcidID{0000-0001-5185-2324}
\and
Joost-Pieter Katoen\inst{1}$^{\text{(\Letter)}}$\orcidID{0000-0002-6143-1926}
\and
Christoph Matheja\inst{3}$^{\text{(\Letter)}}$\orcidID{0000-0001-9151-0441}
\and
Philipp Schr\"oer\inst{1}\orcidID{0000-0002-4329-530X}
}

\institute{ RWTH Aachen University, Aachen, Germany\\
\email{\{kevin.batz,chenms,katoen\}@cs.rwth-aachen.de}
\and
University College London, London, United Kingdom\\
\email{b.kaminski@ucl.ac.uk}
\and
ETH Z\"urich, Z\"urich, Switzerland\\
\email{cmatheja@inf.ethz.ch}
 }


\maketitle


\setlength{\floatsep}{1\baselineskip}
\setlength{\textfloatsep}{1\baselineskip}
\setlength{\intextsep}{1\baselineskip}

\setcounter{footnote}{0}

\begin{abstract}

We revisit two well-established verification techniques, \emph{$k$-in{\-}duc{\-}tion} and \emph{bounded model checking} (BMC), in the more general setting of fixed point theory over complete lattices.
Our main theoretical contribution is \emph{latticed $k$-induction}, which (i) generalizes classical $k$-induction for verifying transition systems, (ii) generalizes Park induction for bounding fixed points of monotonic maps on complete lattices, and (iii) extends from naturals $k$ to transfinite ordinals $\lak$, thus yielding \emph{$\lak$-induction}.

The lattice-theoretic understanding 
of $k$-induction and BMC enables us to apply both techniques to the \emph{fully automatic verification of infinite-state probabilistic programs}. 
Our prototypical implementation manages to automatically verify non-trivial specifications for probabilistic programs taken from the literature that---using existing techniques---cannot be verified without synthesizing a stronger inductive invariant first.

%


\end{abstract}

\lstset{
	basicstyle=\ttfamily,
	keywords=[3]{while,if,else,nat},
	tabsize=2,
	breaklines=true
}


\section{Introduction}

Bounded model checking (BMC)~\cite{DBLP:conf/tacas/BiereCCZ99,DBLP:journals/fmsd/ClarkeBRZ01} is a successful method for analyzing models of hardware and software systems. 
For checking a \emph{finite-state} transition system (TS)
against a safety property (\enquote{bad states are unreachable}), 
BMC unrolls the transition relation until it either finds a counterexample and hence refutes the property, or reaches a pre-computed completeness threshold on the unrolling depth and accepts the property as verified.
For \emph{infinite-state} systems, however, 
such completeness thresholds need not exist (cf.~\cite{DBLP:journals/fmsd/SchuleS07}), rendering BMC a \emph{refutation-only} technique.
To \emph{verify} infinite-state systems, BMC is typically combined with the search for an \emph{inductive invariant}, i.e., a superset of the reachable states which is closed under the transition relation. 
Proving a---not necessarily inductive---safety property then amounts to \emph{synthesizing} a sufficiently strong, often complicated, inductive invariant that excludes the bad states.
A plethora of techniques target 
computing or approximating inductive invariants, including IC3~\cite{DBLP:conf/vmcai/Bradley11}, induction~\cite{DBLP:conf/charme/DeharbeM97,BiereCAV99}, interpolation~\cite{DBLP:conf/cav/McMillan03,DBLP:journals/tcs/McMillan05}, and predicate abstraction~\cite{DBLP:conf/cav/GrafS97,DBLP:conf/tacas/JhalaM06}. 
However, invariant synthesis may burden full automation, 
as it either relies on user-supplied annotations
or confines push-button technologies to semi-decision or approximate procedures.


\emph{$k$-induction}~\cite{k_induction_seminal} 
generalizes the principle of simple induction (aka 1-induction) 
by considering $k$ consecutive transition steps instead of only a single one.
It is more powerful: 
an invariant can be $k$-inductive for some $k{>}1$ but not 1-inductive. 
Following the seminal work of Sheeran et al.~\cite{k_induction_seminal} which combines $k$-induction with SAT solving to check safety properties, $k$-induction has found a broad spectrum of applications in the realm of hardware~\cite{k_induction_seminal,property_directed_k_induction,k_induction_without_unrolling,interpolating_strong_induction} and software verification~\cite{sofware_k_induction,k_induction_multicore_processor,k_induction_dma_races,k_induction_used_by_depthk,boosting_k_induction,kind_cav3}. 
Its success is due to (1) being a foundational yet potent reasoning technique, and
(2) integrating well with SAT/SMT solvers, 
as also pointed out in~\cite{interpolating_strong_induction}: 
\enquote{\emph{the simplicity of applying $k$-induction made it the go-to technique for SMT-based infinite-state model checking}}.
This paper explores whether $k$-induction can have a similar impact on the \emph{fully automatic 
verification} of infinite-state \emph{probabilistic programs}.
That is, we aim to verify that
the \emph{expected value} of a specified \emph{quantity}---think: \enquote{quantitative postcondition}---after the execution of a probabilistic program 
is bounded by a specified threshold.%
%
%
%
%
\begin{example}[Bounded Retransmission Protocol~\textnormal{\cite{DBLP:conf/types/HelminkSV93,DBLP:conf/papm/DArgenioJJL01}}]
\label{ex:brp}
	The loop%
	\begin{align*}
		&\WHILE{\varsent < \vartosend \wwedge \varfailed < \varmaxfailed} 	 \\
		& \hspace{1em} \PCHOICE{\COMPOSE{\ASSIGN{\varfailed}{0}}{
			\ASSIGN{\varsent}{\varsent + 1}
		} }
		%
		{0.9} 
		%
		{\COMPOSE{\ASSIGN{\varfailed}{\varfailed + 1}}{\ASSIGN{\vartotalfailed}{\vartotalfailed + 1}}}  \\
		&\}	
	\end{align*}
	models a simplified version of the
    bounded retransmission protocol, which attempts to transmit $\vartosend$ packages via an unreliable channel (that fails with probability $0.1$) allowing for at most $\varmaxfailed$ retransmissions per package.%
    
    Using our generalization of $k$-induction, we can fully automatically verify that the \emph{expected total number of failed transmissions} is at most $1$,
    if the number of packages we want to (successfully) send is at most $3$.
    In terms of weakest preexpectations~\cite{Kozen1985,McIverM05,benni_diss}, this quantitative property reads
\begin{align*}
	&\wp{\cc}{\vartotalfailed} 
	\ppreceq
	\iverson{\vartosend \leq 3} \cdot (\vartotalfailed + 1) + \iverson{\vartosend > 3}\cdot \infty~.
\end{align*}
The bound on the right-hand-side of the inequality is $4$-inductive, but \emph{not} $1$-inductive; verifying the same bound using $1$-induction requires finding a non-trivial---and far less perspicuous---inductive invariant.
Moreover, if we consider an arbitrary number of packages to send, i.e., we drop $\iverson{\vartosend \leq 3}$, this bound becomes invalid.
In this case, our BMC procedure produces a counterexample, i.e., values for $\vartosend$ and $\varmaxfailed$, proving that the bound does not hold.
\exped
\end{example}%
%
Lifting the classical formalization (and SAT encoding) of $k$-induction over TSs to the probabilistic setting is non-trivial. We encounter the following challenges:

(A) \emph{Quantitative reachability.}
In a TS, a state reachable within $k$ steps remains reachable on increasing $k$.
In contrast, reachability \emph{probabilities} in Markov chains---a common operational model for probabilistic programs~\cite{DBLP:journals/pe/GretzKM14}---may increase on increasing $k$.
Hence, proving that the probability of reaching a bad state remains below a given threshold is more intricate than reasoning about qualitative reachability.

(B) \emph{Counterexamples are subsystems.}
In a TS, an acyclic path from an initial to a bad state suffices as a witness for refuting safety, i.e., non-reachability.
SAT encodings of $k$-induction rely on this by expressing the absence of witnesses up to a certain path-length.
In the probabilistic setting, however, witnesses are no longer single paths~\cite{DBLP:journals/tse/HanKD09}.
Rather, a witness for the probability of reaching a bad state to exceed a threshold is a \emph{subsystem}~\cite{DBLP:journals/tocl/ChadhaV10}, i.e., a set of possibly cyclic paths.

(C) \emph{Symbolic encodings.}
To enable fully automated verification, we need a suitable encoding
such that our lifting integrates well into SMT solvers.
Verifying probabilistic programs involves reasoning about execution \emph{trees}, where each (weighted) branch corresponds to a probabilistic choice. 
A suitable encoding needs to capture such trees which
requires more involved theories than encoding paths in classical $k$-induction.

We address challenges (A) and (B) by developing \emph{latticed $k$-induction}, which is a proof technique in the rather general setting of fixed point theory over complete lattices. 
Latticed $k$-induction generalizes classical $k$-induction in three aspects: 
(1) it works with any monotonic map on a complete lattice instead of being confined to the transition relation of a 
transition system, (2)~it generalizes the Park induction principle for bounding fixed points of such monotonic maps, and (3) it extends from natural numbers $k$ to (possibly transfinite) ordinals $\lak$, hence its short name: \emph{$\lak$-induction}.

It is this lattice-theoretic understanding that enables us to lift both $k$-in{\-}duc{\-}tion and BMC to reasoning about quantitative properties of probabilistic programs.
To enable \emph{automated} reasoning, we address challenge (C) by an incremental SMT encoding 
based on the theory of quantifier-free mixed integer and real arithmetic with uninterpreted functions ($\qfuflira$).
We show how to effectively compute all needed operations for $\lak$-induction using the SMT encoding and, in particular, how to decide \emph{quantitative entailments}.

A prototypical implementation of our method demonstrates that $\lak$-induction for (linear) probabilistic programs manages to automatically verify non-trivial specifications for programs taken from the literature which---using existing tech\-ni\-ques---cannot be verified without synthesizing a stronger inductive invariant.
%
	%
%

\paragraph{\bfseries \upshape{Related Work.}} 
Besides the aforementioned related work on $k$-induction, we briefly discuss other
automated analysis techniques for probabilistic systems 
and
other approaches for bounding fixed points.
%
Symbolic engines exist for exact inference~\cite{DBLP:conf/cav/GehrMV16} and sensitivity analysis~\cite{DBLP:conf/atva/HuangWM18}. 
Other automated approaches focus on bounding expected costs~\cite{DBLP:conf/pldi/NgoC018}, termination analysis~\cite{DBLP:conf/cav/ChakarovS13,DBLP:journals/pacmpl/AgrawalC018}, and static analysis~\cite{DBLP:conf/pldi/WangHR18,DBLP:journals/toplas/AmtoftB20}.
BMC has been applied in a rather rudimentary form to the on-the-fly verification of finite unfoldings of probabilistic programs~\cite{DBLP:conf/atva/0001DKKW16}, and the enumerative generation of counterexamples in finite Markov chains~\cite{DBLP:conf/vmcai/WimmerBB09}.
\mbox{(Semi-)auto}\-mated invariant-synthesis techniques can be found in~\cite{DBLP:conf/sas/KatoenMMM10,DBLP:conf/atva/FengZJZX17,DBLP:conf/cav/BartheEFH16}.
A recent variant of IC3 for probabilistic programs called PrIC3~\cite{pric3} 
is restricted to finite-state systems.
When applied to finite-state Markov chains,
our $\lak$-induction operator is related to other operators that
have been employed for determining reachabilitiy probabilities
through value iteration~\cite{DBLP:conf/cav/QuatmannK18,DBLP:conf/cav/Baier0L0W17,DBLP:conf/cav/HartmannsK20}.
In particular, when iterated on the candidate upper bound, the $\lak$-induction operator
coincides with the (upper value iteration) operator in interval iteration~\cite{DBLP:conf/cav/Baier0L0W17};
the latter operator can be used together with the
up-to techniques (cf.~\cite{DBLP:books/daglib/0067019,DBLP:conf/aplas/Pous07,DBLP:books/cu/12/PousS12})
to prove our $\lak$-induction rule sound (in contrast, we give an elementary proof).
However, the $\lak$-induction operator avoids comparing current and previous iterations. It is thus easier to implement and more amenable to SMT solvers.
%
Finally, the proof rules for bounding fixed points recently developed in~\cite{DBLP:journals/corr/abs-2101-08184}
are restricted to finite-state systems. 

\section{Verification as a Fixed Point Problem}
\label{sec:problem_statement}

%
We start by recapping some fundamentals on fixed points of monotonic operators on complete lattices
before we state our target verification problem. 

\paragraph{Fundamentals.}
For the next three sections, we fix a \emph{complete lattice} $(\laset,\, \laord)$, i.e.\ a carrier set $E$ together with a partial order $\sqsubseteq$, such that every subset $S \subseteq E$ has a \emph{greatest lower bound}~$\bigsqcap S$ (also called the \emph{meet} of $S$) and a \emph{least upper bound}~$\bigsqcup S$ (also called the \emph{join} of $S$).
For just two elements $\{\lag, \lah\} \subseteq \laset$, we denote their meet by $\lag \sqcap \lah$ and their join by $\lag \sqcup \lah$.
Every complete lattice has a \emph{least} and a \emph{greatest} element, which we denote by $\bot$ and $\top$, respectively.

In addition to $(\laset,\, \laord)$, we also fix a \emph{monotonic operator} $\laop \colon \laset \to \laset$.
By the Knaster-Tarski theorem~\cite{knaster1928theoreme,tarski1955lattice,DBLP:journals/ipl/LassezNS82}, every monotonic operator~$\laop$ admits a \emph{complete lattice of (potentially infinitely many) fixed points}.
The least fixed point~$\lfp \laop$ and the greatest fixed point~$\gfp \laop$ are moreover constructible by (possibly transfinite) \emph{fixed point iteration} from~$\bot$ and~$\top$, respectively:
Cousot \& Cousot~\cite{cousot1979constructive}
showed that there exist ordinals~$\oa$ and $\ob$, such that\footnote{We use lowercase greek letters $\oa$, $\ob$, $\oc$, $\od$, etc. to denote arbitrary (possibly transfinite) ordinals and $i$, $j$, $k$, $m$, $n$, etc. to denote natural (finite) numbers in $\Nats$.}
\begin{align}\label{eq:fixedpoints}
    \lfp \laop \eeq \lappup{\alpha}{\bot} \qand \gfp \laop \eeq \lappdown{\beta}{\top}~, \tag{$\dagger$}
\end{align}
where $\lappup{\od}{\lag}$ denotes the \emph{upper $\od$-fold iteration}
and $\lappdown{\od}{\lag}$ denotes the \emph{lower $\od$-fold iteration} 
of $\laop$ on $\lag$, respectively.
Formally, $\lappup{\od}{\lag}$ is given by%
\begin{align*}
    \lappup{\od}{\lag} \eeq 
    \begin{cases}
        \lag & ~\text{if}~ \od = 0~, \\[.5em]
        \lapp{\lappup{\oc}{\lag}} & ~\text{if}~ \od = \oc + 1 ~\text{is a successor ordinal}~, \\[.5em]
        \bigsqcup \left\{ \lappup{\oc}{\lag} \mathrel{\big|} \oc < \od \right\} &  ~\text{if}~ \od ~\text{is a limit ordinal\footnotemark}~.
    \end{cases}
\end{align*}%
\footnotetext{%
\label{footnote}%
To ensure well-definedness of transfinite iterations, 
we fix an \emph{ambient ordinal} $\lambient$ and \emph{tacitly assume $\od < \lambient$ for all ordinals $\od$ 
considered throughout this paper.}
Formally, $\lambient$ is the smallest ordinal such that $|\lambient| > |\laset|$. 
Intuitively, $\lambient$~then upper-bounds the length of any repetition-free sequence over elements of $\laset$.%
}%
Intuitively, if $\od$ is the successor of $\oc$, then we simply do another iteration of~$\Phi$.
If $\od$ is a limit ordinal, then $\lappup{\od}{\lag}$ can also be thought of as a limit, namely of iterating $\laop$ on $\lag$.
However, simply iterating $\laop$ on $\lag$ need not always converge, especially if the iteration does not yield an ascending chain.
To remedy this, we take as limit the join over the whole (possibly transfinite) iteration sequence, i.e., the least upper bound over all elements that occur along the iteration.
The lower $\od$-fold iteration $\lappdown{\od}{\lag}$ is defined analogously to $\lappup{\od}{\lag}$, except that we take a meet instead of a join whenever $\od$ is a limit ordinal.

An important special case for fixed point iteration (see~\eqref{eq:fixedpoints}) is when the operator~$\laop$ is \emph{Scott-continuous} (or simply \emph{continuous}), i.e.,\ if $\lapp{\bigsqcup\vphantom{\bigl(} \{\lag_1 \laord \lag_2 \laord {\ldots}\}} = \bigsqcup \lapp{\vphantom{\bigl(}\{\lag_1 \laord \lag_2 \laord {\ldots}\}\vphantom{\bigl(}}$.
In this case, $\oa$ in~\eqref{eq:fixedpoints} coincides with the first infinite limit ordinal $\omega$ (which can be identified with the set $\Nats$ of natural numbers).
This fact is also known as the Kleene fixed point theorem~\cite{Abramsky94domaintheory}.

\paragraph{Problem statement.}

Fixed points are ubiquitous in computer science.
Prime examples of properties that can be conveniently characterized as least fixed points include both the set of reachable states in a transition system
and the function mapping each state in a Markov chain to the probability of reaching some goal state (cf.~\cite{Puterman1994Markov}).
However, least and greatest fixed points are often difficult or even impossible~\cite{%
DBLP:journals/acta/KaminskiKM19} to compute; it is thus desirable to \emph{bound} them.

For example, it may be sufficient to prove that a system modeled as a Markov chain reaches a bad state from its initial state with probability \emph{at most} $10^{{-}6}$, instead of computing \emph{precise} reachability probabilities for each state.
Moreover, if said probability is \emph{not} bounded by $10^{{-}6}$, we would like to witness that as well.
%

%
In general lattice-theoretic terms, our problem statement reads as follows:%
\begin{center}\fbox{%
	\parbox{.98\textwidth}{%
		\centering{
			\vspace{.4em}
			Given a complete lattice $(\laset,\, {\laord})$, a monotonic operator~$\laop\colon \laset \to \laset$,\linebreak and a candidate upper bound $\laf \in \laset$ on $\lfp \laop$, %
			\abovedisplayskip=.5em%
			\begin{align*}
				\textnormal{\emph{prove} or \emph{refute} that } \lfp \laop \llaord \laf~.
			\end{align*}%
			\vspace*{-1.4em}
		}%
	}%
}%
\end{center}
For \emph{proving}, we will present \emph{latticed $k$-induction};
for \emph{refuting}, we will present \emph{latticed bounded model checking}. 
Running both in parallel may (and under certain conditions: \emph{will}) lead to a decision of the above problem.

\section{Latticed $k$-Induction}
\label{sec:k_induction}

%
In this section, we generalize the well-established $k$-induction verification technique~\cite{k_induction_seminal,sofware_k_induction,kind_cav3,property_directed_k_induction,k_induction_without_unrolling,interpolating_strong_induction} to \emph{latticed $k$-induction} (for short: \emph{$\lak$-induction}; reads: \enquote{kappa induction}).
With $\lak$-induction, our aim is to \emph{prove} that $\lfp \laop \laord \laf$.
To this end, we attempt \enquote{ordinary} induction, also known as \emph{Park induction}:%
\begin{theorem}[Park Induction~\textnormal{\cite{park1969fixpoint}}]
\label{thm:park}
	Let $\laf \in \laset$.
	Then%
	\begin{align*}
        \lapp{\laf} \llaord \laf \qimplies \lfp \laop \llaord \laf~.
	\end{align*}%
\end{theorem}%
Intuitively, this principle says: 
if pushing our candidate upper bound $\laf$ through~$\laop$ takes us \emph{down} in the partial order $\laord$, we have verified that $\laf$ is indeed an upper bound on $\lfp \laop$.
The true power of Park induction is that applying $\laop$ \emph{once} tells us something about iterating~$\laop$ possibly \emph{transfinitely often} (see~\eqref{eq:fixedpoints} in \Cref{sec:problem_statement}).

Park induction, unfortunately, does \emph{not} work in the reverse direction:
If we are unlucky, $\laf \sqsupsetneq \lfp \laop$ \emph{is} an upper bound on $\lfp \laop$, but nevertheless $\lapp{\laf} \not\laord \laf$.
In this case, we say that $\laf$ is \emph{not inductive}.
But how can we verify that $\laf$ is indeed an upper bound in such a non-inductive scenario?
%
We search \emph{below~$\laf$} for a \emph{different, but inductive}, upper bound on $\lfp \Phi$, that is, we 
%
\begin{align*}
    \textnormal{search for an } \lah \in \laset ~\quad \textnormal{ such that } \quad~ \lfp \laop \llaord \laop(\lah) \llaord \lah \llaord \laf~.
\end{align*}%
In order to perform a \emph{guided} search for such an $\lah$, we introduce the \lak-induction operator---a modified version of $\laop$ that is parameterized by our candidate $\laf$:%
\begin{definition}[$\boldsymbol{\lak}$-Induction Operator]
	\label{def:kindop}
	For $\laf \in \laset$, we call
	\[
        \kindop{\laf} \colon\quad \laset \tto \laset, \qquad \lag \mmapsto \lapp{\lag} \sqcap \laf
	\]
	the \emph{\lak-induction operator} (with respect to $\laf$ and $\laop$). 
\end{definition}
What does $\kindop{\laf}$ do?
As illustrated in 
\Cref{fig:kappa-induction},
\begin{figure}[t]
	\centering
	\begin{adjustbox}{max width=.85\linewidth}
	\scalebox{1.4}{
	\begin{tikzpicture}[dot/.style={draw,circle,minimum size=1mm,inner sep=0pt,outer sep=2pt,fill=black},
		cross/.style={cross out, draw=black, minimum size=2*(#1-\pgflinewidth), inner sep=0pt, outer sep=0pt},
		cross/.default={1pt}
		]
		
		
		
		\node (ghost1) at (1.4,0.3) {};
		\draw [RedViolet,opacity=.5] (-0.325,-0.325) rectangle (3.625,1.18) {};
		\node[RedViolet,below right] at (ghost1.south east) {\scriptsize Latticed BMC};
		
		\node (ghost2) at (8.25,0.5) {};
		\draw [NavyBlue,opacity=.5] (2.83,0.5) rectangle (8.25,4.5) {};
		\node[NavyBlue,above left] at (ghost2.north west) {\scriptsize $\lak$-induction};
		
		\node () at (-0.12,-0.12) {\tiny $\bot$};
%
		
		\node [dot] (bot0) at (0,0) {};
		\node [dot,label={[xshift=-.03cm,yshift=-.11cm]{\tiny $\laop^{\scaleto{\left\lceil 1 \right\rceil}{3.5pt}}(\bot)$}}] (bot1) at (0.5,0.5) {};
		\node [dot,label={[xshift=.03cm,yshift=-.6cm]{\tiny $\laop^{\scaleto{\left\lceil 2 \right\rceil}{3.5pt}}(\bot)$}}] (bot2) at (1.25,0.75) {};
		\node [dot,label={[xshift=.03cm,yshift=-.6cm]{\tiny $\laop^{\scaleto{\left\lceil \od \right\rceil}{3.5pt}}(\bot)$}}] (botdelta) at (2,1) {};
		
		\draw[lightgray] 
		(2.9,1.9) .. controls (2.4,2.05) and (2.4,2.3) .. 
		(2.7,2.4) .. controls (2.5,2.7) and (2.5,3.4) .. 
		(2.9,3.1) .. controls (3.1,3.3) and (3.2,3.3) .. 
		(3.5,3.1) .. controls (3.4,3.3) and (3.6,3.4) .. 
		(3.85,3.1) .. controls (4.1,3.0) and (4.2,2.8) .. 
		(3.95,2.7) .. controls (4.3,2.4) and (4.1,2.2) .. 
		(3.85,1.9) .. controls (3.6,1.8) and (3.5,1.75) .. 
		(3.4,1.9) .. controls (3.2,1.8) and (3.0,1.75) .. 
		cycle;
		
		\node [dot,NavyBlue,label={[yshift=-.6cm,xshift=.27cm]{\textcolor{NavyBlue}{\scriptsize $\lfp\! \laop$}}}] (lfp) at (3,1) {};
		\node [dot,lightgray] (fp) at (3.75,2.0) {};
		\node [dot,lightgray,label={[xshift=-.65cm,yshift=.15cm,lightgray]{\tiny $\textsf{fp's}\: \laop$}}] (fp2) at (3,2.0) {};
		\node [dot,lightgray] (fp3) at (3.75,3.0) {};
		\node [dot,lightgray] (fp4) at (3,3.0) {};
		\node [dot,lightgray,label={[yshift=-.1cm,lightgray]{\tiny $\gfp\! \laop$}}] (gfp) at (3.75,4) {};
		
		\draw[->,>=stealth]
		(bot0) edge (bot1)
		(bot1) edge (bot2)
		(bot2) edge[dotted] (botdelta)
		(botdelta) edge (lfp)
		(lfp) edge[lightgray] (fp)
		(lfp) edge[lightgray] (fp2)
		(fp3) edge[lightgray] (gfp)
		(fp4) edge[lightgray] (gfp);
		
		\draw[->,>=stealth,dotted,lightgray]
		(fp) edge (fp3)
		(fp) edge (fp4)
		(fp2) edge (fp3)
		(fp2) edge (fp4);
		
		
		\node [dot,label={[xshift=-.48cm,yshift=-.35cm]{\tiny $\Psi_{\scaleto{\laf}{3.5pt}}^{\scaleto{\left\lfloor \lak \right\rfloor}{3.5pt}}(f)$}}] (psik) at (5.75,2.3) {};
		\node [dot,label={[xshift=-.48cm,yshift=-.4cm]{\tiny $\Psi_{\scaleto{\laf}{3.5pt}}^{\scaleto{\left\lfloor 2 \right\rfloor}{3.5pt}}(f)$}}] (psi2) at (5.75,2.9) {};
		\node [dot,label={[xshift=-.48cm,yshift=-.4cm]{\tiny $\Psi_{\scaleto{\laf}{3.5pt}}^{\scaleto{\left\lfloor 1 \right\rfloor}{3.5pt}}(f)$}}] (psi1) at (5.75,3.5) {};
		\node [dot,NavyBlue,label={[yshift=-.1cm]{\textcolor{NavyBlue}{\scriptsize $f$}}}] (f) at (5.75,4) {};
		
		\draw[->,>=stealth]
		(psik) edge[dotted] (psi2)
		(psi2) edge (psi1)
		(psi1) edge (f);
		
		\node [dot,label={[xshift=.12cm,yshift=-.65cm]{\tiny $\laop(\Psi_{\scaleto{\laf}{3.5pt}}^{\scaleto{\left\lfloor \lak \right\rfloor}{3.5pt}}(f)$)}}] (phipsik) at (5,1.5) {};
		\node [dot,label={[xshift=.8cm,yshift=-.4cm]{\tiny $\laop(\Psi_{\scaleto{\laf}{3.5pt}}^{\scaleto{\left\lfloor \lak-1 \right\rfloor}{3.5pt}}(f))$}}] (phipsi2) at (6.25,2.3) {};
		\node [dot,label={[xshift=.7cm,yshift=-.4cm]{\tiny $\laop(\Psi_{\scaleto{\laf}{3.5pt}}^{\scaleto{\left\lfloor 1 \right\rfloor}{3.5pt}}(f))$}}] (phipsi1) at (6.75,2.9) {};
		\node [dot,label={[yshift=-.1cm]{\tiny $\laop(f)$}}] (phif) at (7.25,3.5) {};
		
		\draw[->,>=stealth]
		(lfp) edge[NavyBlue,dashed,thick] (phipsik)
		(phipsik) edge[NavyBlue,thick,dashdotted] (psik)
		(phipsi2) edge[dotted] (phipsi1)
		(phipsi1) edge (phif)
		
		(phipsik) edge (phipsi2)
		(psi1) edge (phif)
		(psi2) edge (phipsi1)
		(psik) edge (phipsi2)
		(phipsik) edge[thick,NavyBlue,bend left=83] (f);

	\end{tikzpicture}
	}
	\end{adjustbox}

	\caption{$\lak$-induction and latticed BMC in case that $\lfp \laop \laord f$. 
	An arrow from~$\lag$ to~$\lah$ indicates $\lag \laord \lah$.
	The solid blue arrow from \smash{$\Phi(\kindopfold{\laf}{\lak}(\laf))$} to $\laf$ is the premise of $\kappa$-induction, i.e., the LHS of \Cref{thm:k_induction_to_induction}, which implies
	the dash-dotted blue arrow from \smash{$\Phi(\kindopfold{\laf}{\lak}(\laf))$} to \smash{$\kindopfold{\laf}{\lak}(\laf)$}, i.e., the RHS of \Cref{thm:k_induction_to_induction}.
	The dashed blue arrow from $\lfp \laop$ to \smash{$\Phi(\kindopfold{\laf}{\lak}(\laf))$} is a consequence of the dash-dotted arrow (by Park induction, \Cref{thm:park}) and ultimately proves that $\lfp \laop \laord f$.
	}
	\label{fig:kappa-induction}
\end{figure}%
%
if $\lapp{\laf} \not\laord \laf$ (i.e.\ $\laf$ is non-inductive) then \enquote{\emph{at least some part of $\lapp{\laf}$ is greater than $\laf$}}.
If the whole of $\lapp{\laf}$ is greater than $\laf$, then $\laf \sqsubsetneq \lapp{\laf}$; 
if only some part of $\lapp{\laf}$ is greater and some is smaller than $\laf$, then $\laf$ and $\lapp{\laf}$ are incomparable.
The \lak-induction operator~$\kindop{\laf}$ now \emph{rectifies} $\lapp{\laf}$ being (partly) greater than $\laf$ by \emph{pulling $\lapp{\laf}$ down} via the meet with $f$ (i.e., via $\cloze{a} \sqcap \laf$), so that the result is in no part greater than $f$.
Applying $\kindop{\laf}$ to $\laf$ hence always yields something below or equal to $f$. 

Together with the observation that~$\kindop{\laf}$ is monotonic, iterating $\kindop{\laf}$ on $\laf$ necessarily \emph{descends} from~$\laf$ downwards in the direction of $\lfp \laop$ (and never below):%
\begin{lemma}[Properties of the $\boldsymbol{\lak}$-Induction Operator]
	\label{lem:kindop_mon_chain}
	Let $\laf \in \laset$ and let~$\kindop{\laf}$ be the \lak-induction operator with respect to $\laf$ and $\laop$.
	Then%
	\begin{enumerate}[(a)]
        \item
		\label{lem:kindop_mon_chain_1} 
			$\kindop{\laf}$ is monotonic, i.e., %
			$
				\forall\, \lag_1,\lag_2 \in \laset \colon~ 
				\lag_1 \llaord \lag_2 
				~~\textnormal{implies}~~
				\kindop{\laf}(\lag_1) \llaord \kindop{\laf}(\lag_2) 
			$.

		\item 
		\label{lem:kindop_mon_chain_2}
            Iterations of $\kindop{\laf}$ starting from $f$ are descending, i.e.,
            for all ordinals
            $\oc$, $\od$,
			%
			\begin{align*}
				\oc \LL \od 
				\qimplies
				\kindopfold{\laf}{\od}(\laf) \llaord \kindopfold{\laf}{\oc}(\laf)~.
			\end{align*}%

		\item 
		\label{lem:kindop_mon_chain_4}
            $\kindop{\laf}$ is dominated by $\laop$, i.e.,
			$\forall\, \lag \in \laset\colon~
                \kindop{\laf}(\lag) \llaord \lapp{\lag}~.
			$
			\medskip
			
		\item 
		\label{lem:kindop_mon_chain_3}
			If $\lfp \laop \laord \laf$, then for any ordinal $\od$,
			\begin{align*}
                \lfp \laop 
                \llaord {\ldots} \llaord 
                \kindopfold{\laf}{\od}(\laf) 
                \llaord {\ldots} \llaord 
                \kindopfold{\laf}{2}(\laf) 
                \llaord \kindop{\laf}(\laf) \llaord f~.
			\end{align*}%
	\end{enumerate}%
\end{lemma}
\begin{proof}
	See Appendix~\ref{proof:lem:kindop_mon_chain}. \qed
\end{proof}
The descending sequence
$\laf \sqsupseteq \kindop{\laf}(\laf) \sqsupseteq \kindopfold{\laf}{2}(\laf) \sqsupseteq {\ldots}$ constitutes our guided search for an inductive upper bound on $\lfp \Phi$.
For each ordinal $\lak$ (hence the short name: \lak-induction), $\smash{\kindopfold{\laf}{\lak}(f)}$ is a potential candidate for Park induction:%
\begin{align}\label{eq:candidateParkInduction}
	\Phi\left(\kindopfold{\laf}{\lak}(f)\right) 
    \smash{~{}\stackrel{\textnormal{\tiny potentially}}{\laord}{}~} 
    \kindopfold{\laf}{\lak}(f)~. \tag{$\ddagger$}
\end{align}
For efficiency reasons, e.g.,\ when offloading the above inequality check to an SMT solver, we will not check the inequality~\eqref{eq:candidateParkInduction} directly but a property equivalent
to~\eqref{eq:candidateParkInduction}, namely whether \smash{$\Phi(\kindopfold{\laf}{\lak}(\laf))$} is below~$\laf$ instead of \smash{$\kindopfold{\laf}{\lak}(\laf)$}:%
\begin{lemma}[Park Induction from $\boldsymbol{\lak}$-Induction]
	\label{thm:k_induction_to_induction}
	Let $\laf \in \laset$. 
    Then 
	\[
       \lapp{ \kindopfold{\laf}{\lak} (\laf) } \llaord \laf 
	   \qiff
       \lapp{ \kindopfold{\laf}{\lak}(\laf) } \llaord \kindopfold{\laf}{\lak}(\laf)~.
	\] 
\end{lemma}
\begin{proof} 
    %
	The if-direction is trivial, as $\kindopfold{\laf}{\lak}(\laf) \laord \laf$ (\Cref{lem:kindop_mon_chain}\ref{lem:kindop_mon_chain_3}). For only-if:
    \begin{align*}
     \kindopfold{\laf}{\lak}(\laf)
    \llaordrev & 
    \kindopfold{\laf}{\lak+1}(\laf) 
    \tag{by \Cref{lem:kindop_mon_chain}\ref{lem:kindop_mon_chain_2}} \\
    \eeq & 
    \kindopapp{\laf}{\kindopfold{\laf}{\lak}(\laf) }  
    \tag{by definition of $\kindopfold{\laf}{\lak+1}(\laf)$} \\
    \eeq &
    \lapp{ \kindopfold{\laf}{\lak}(\laf) } \sqcap \laf
    \tag{by definition of $\kindop{\laf}$} 
    \\
    \llaordrev & 
    \lapp{ \kindopfold{\laf}{\lak}(\laf) }~.
        \tag*{(by the premise)~\qed}
    \end{align*}
\end{proof}%
If $\Phi \bigl( \kindopfold{\laf}{\lak} (\laf) \bigr) \laord \laf$, then \Cref{thm:k_induction_to_induction} tells us that $\kindopfold{\laf}{\lak} (\laf)$ is Park inductive and thereby an upper bound on $\lfp \Phi$.
Since iterating $\kindop{\laf}$ on $f$ yields a descending iteration sequence (see \Cref{lem:kindop_mon_chain}\ref{lem:kindop_mon_chain_2}), \smash{$\kindopfold{\laf}{k} (\laf)$}~is below $f$ and therefore $f$ is also an upper bound on $\lfp \Phi$.
Put in more traditional terms, we have shown that \smash{$\kindopfold{\laf}{\lak}(\laf)$} is an inductive invariant stronger than $\laf$.
Formulated as a proof rule, we obtain the following induction principle:
\begin{theorem}[$\boldsymbol{\lak}$-Induction]
	\label{cor:k_induction}
	Let $\laf \in E$ and let $\lak$ be an ordinal. Then
	\[
	\Phi\left(\kindopfold{\laf}{\lak}(\laf) \right) \llaord \laf  
	\qimplies
	\lfp \Phi \llaord \laf~.
	\] 
\end{theorem}
\begin{proof}
	Following the argument above, for details see \Cref{proof:cor:k_induction}.\qed
\end{proof}
An illustration of $\lak$-induction is shown in (the right frame of) \Cref{fig:kappa-induction}. 
For every ordinal~$\lak$, if~\smash{$\Phi(\kindopfold{\laf}{\lak}(\laf))\laord \laf$}, 
then we call 
$\laf$~\emph{$(\lak{+}1)$-inductive} (for~$\laop$).
In particular, $\lak$-induction generalizes Park induction, in the sense that $1$-induction \emph{is} Park induction and, 
$(\lak{>}1)$-induction is a \emph{more general principle of induction}.

\begin{figure}[!t]
\begin{minipage}[t]{0.49\textwidth}
 \begin{algorithm}[H]
     \caption{\mbox{Latticed $k$-induction}}
		\label{alg:k_induction}
		\SetKwInput{Input}{input}\SetKwInOut{Output}{output}\SetNoFillComment
		\Input{$\laop\colon \laset \to \laset$ and $\laf \in \laset$.}
        \Output{\enquote{$\verify$} if $\laf$ is a $k$-inductive invariant, diverge otherwise.}
		%
		$\lag \leftarrow \laf$ \;
        \While{$\lapp{\lag} \not\laord \laf$\label{line_loop-alg:k_induction}}{
            $\lag \leftarrow \kindop{\laf}(\lag)~$\tcp*{$\texttt{recall: } \kindop{\laf}(\lag) = \lapp{\lag} \sqcap \laf$}\label{line_body-alg:k_induction}
		}
		\Return $\verify$ \;
	\end{algorithm}
\end{minipage}
\hfill
\begin{minipage}[t]{0.49\textwidth}
    \begin{algorithm}[H]
        \caption{Latticed BMC}
		\label{alg:bmc}
		\SetKwInput{Input}{input}\SetKwInOut{Output}{output}\SetNoFillComment
		\Input{$\laop\colon \laset \to \laset$ and $\laf \in \laset$.}
        \Output{\enquote{$\refute$} if there exists $k \in \Nats$ with $\lappup{k}{\bot} \not\laord \laf$, diverge otherwise.}
		$\lag \leftarrow \bot$ \;
		\Repeat{$\lag \not\laord \laf~$}{
            $\lag \leftarrow \lapp{\lag}$ \;
		}
		\vspace*{.135cm}
		\Return $\refute$ \;
	\end{algorithm}
\end{minipage}
\end{figure}

Algorithm~\ref{alg:k_induction} depicts a (semi-)algorithm that performs \emph{latticed $k$-induction} (for $k < \omega$) in order to prove~$\lfp \laop \laord \laf$ by iteratively increasing $k$.
For implementing this algorithm, we require, of course, that both $\laop$ and $\kindop{\laf}$ are computable and that $\laord$ is decidable.
Notice that the loop (lines~\ref{line_loop-alg:k_induction}--\ref{line_body-alg:k_induction}) never terminates if $\laf \sqsubset \lapp{\laf}$---a condition that can easily be checked before entering the loop. 
Even with this optimization, however, Algorithm~\ref{alg:k_induction} is a \emph{proper} semi-algorithm: even if $\lfp \laop \laord \laf$, then $\laf$ is still not guaranteed to be $k$-inductive for some $k < \omega$.
And even if an algorithm \emph{could} somehow perform transfinitely many iterations, then $\laf$ is still not guaranteed to be $\lak$-inductive for some ordinal $\lak$:%

\begin{counterexample}[Incompleteness of $\boldsymbol{\lak}$-Induction]%
	Consider the carrier set~$\{0, 1, 2\}$, partial order $0 \sqsubset 1 \sqsubset 2$, and the monotonic operator $\laop$ with $\laop(0) = 0 = \lfp \laop$, and $\laop(1) = 2$, and $\laop(2) = 2 = \gfp \laop$.
	Then $\lfp \Phi \laord 1$, but for any ordinal $\lak$, $\kindopfold{1}{\lak}(1) = 1$ and $\laop(1) = 2 \not\laord 1$. Hence $1$ is not $\lak$-inductive.
	\exped
\end{counterexample}%
Despite its incompleteness, we now provide a \emph{sufficient} criterion which ensures that \emph{every} upper bound on $\lfp \laop$ is $\lak$-inductive for some ordinal~$\lak$.
%
\begin{theorem}[Completeness of $\boldsymbol{\lak}$-Induction for Unique Fixed Point]
\label{thm:kind-complete-for-unique}
    If $\lfp \laop = \gfp \laop$ (i.e.\ $\laop$ has \emph{exactly one} fixed point), then, for every $\laf \in \laset$,
	\begin{align*}
        \lfp \Phi \llaord \laf \qimplies \textnormal{$\laf$ is $\lak$-inductive for some ordinal $\lak$}~. 
	\end{align*}
\end{theorem}%
\begin{proof}
	By the Knaster-Tarski theorem, we have $\ladown{\ob}(\top) = \gfp \laop$ for some ordinal~$\ob$.
	We then show that $\laf$ is $(\ob{+}1)$-inductive, see \Cref{proof:thm:kind-complete-for-unique}. \qed
\end{proof}%
The proof of the above theorem immediately yields that, if the unique fixed point can be reached through \emph{finite} fixed point iterations starting at $\top$, then $\laf$ is $k$-inductive for some \emph{natural} number $k$; 
Algorithm~\ref{alg:k_induction} thus eventually terminates.%
\begin{corollary}
    If $\lappdown{n}{\top} = \lfp \laop$ for some $n \in \Nats$, then, for every $\laf \in \laset$, 
	\begin{align*}
        \lfp \Phi 
        \llaord 
        \laf \qimplies \laf \textnormal{ is $n$-inductive for some $n \in \Nats$}~. 
	\end{align*}%
\end{corollary}%

\section{Latticed vs.~Classical $k$-Induction}
\label{sec:latticed-vs-classical}


We show that our purely lattice-theoretic $\lak$-induction from \Cref{sec:k_induction} generalizes classical $k$-induction for hardware- and software verification.
To this end, we first recap how $k$-induction is typically formalized in the literature~\cite{sofware_k_induction,boosting_k_induction,property_directed_k_induction,k_induction_without_unrolling}:
Let $\transsys = (\stateset,\, \init, \transrel)$ be a transition system, where $\stateset$ is a (countable) set of \emph{states}, $\init \subseteq \stateset$ is a non-empty set of \emph{initial states}, and $\transrel \subseteq \stateset \times \stateset$ is a \emph{transition relation}. 
%
%
	As in the seminal work on $k$-induction~\cite{k_induction_seminal}, we require that
    $\transrel$ is a \emph{total} relation, i.e., every state has at least one successor.
    This requirement is sometimes overlooked in the literature, which renders the classical SAT-based formulation of $k$-induction ((\ref{form:kind_init}) and (\ref{form:kind_step}) below) unsound in general.
%

%
Our goal is to verify that a given \emph{invariant property} $\propset \subseteq \stateset$ covers all states reachable in $\transsys$ 
from some initial state.
Suppose that $\init$, $\transrel$ and $\propset$ are characterized by 
logical formulae $\init(s)$, $\transrelf(s,s')$ and $\propsetf(s)$ (over the free variables $s$ and $s'$), respectively. 
Then, achieving the above goal with classical $k$-induction 
amounts to proving the validity of 
{
\begin{subequations}
\begin{align}
	&
    \initf(s_1) \wedge \transrelf(s_1,s_2) \wedge \ldots \wedge \transrelf(s_{k-1},s_k) \implies \propsetf(s_1) \wedge \ldots \wedge \propsetf(s_k)~,\qand
    \label{form:kind_init}
    \\
	&
    \propsetf(s_1) \wedge \transrelf(s_1,s_2)
	\wedge \ldots \wedge \propset(s_k)
    \wedge \transrelf(s_k,s_{k+1})
	\implies \propsetf(s_{k+1})~.
    \label{form:kind_step}
\end{align}%
\end{subequations}%
}%
Here, the \emph{base case}~\eqref{form:kind_init} asserts that $\propset$ holds for \emph{all states reachable within $k$~transition steps from some initial state};
the \emph{induction step}~\eqref{form:kind_step} formalizes that $\propset$ is \emph{closed under taking up to $k$ transition steps}, i.e., if we start in $\propset$ and stay in $\propset$ for up to $k$ steps, then we also end up in $\propset$ after taking the $(k{+}1)$-st step. 
If both~\eqref{form:kind_init} and~\eqref{form:kind_step} are valid, then classical $k$-induction tells us that the property $\propset$ holds for \emph{all} reachable states of $\transsys$.
How is the above principle reflected in \emph{latticed} $k$-induction (cf.~\Cref{sec:k_induction})?
For that, we 
choose the complete lattice $(2^\stateset,\, \subseteq)$, where $2^S$ denotes the powerset of $\stateset$; the least element is $\bot = \emptyset$ and the meet operation is standard intersection $\cap$.

Moreover, we define a monotonic operator $\laop$ 
whose least fixed point precisely characterizes the set of reachable states of the transition system $\transsys$: 
%
\begin{align*}
    \laop\colon\quad 2^\stateset \tto 2^\stateset, \qquad F \mmapsto I \,\cup\, \succs(F)~,
\end{align*}%
That is, $\laop$
maps any given set of states $F \subseteq \stateset$ to the union of the initial states $I$ and of those states $\succs(F)$ that are reachable from $F$ using a single transition.\footnote{Formally, $\succs(F) \ddefeq \setcomp{ t' }{ t \in F,\, (t,t') \in \transrel }$.}

Using the $\lak$-induction operator $\kindop{\propset}$ constructed from $\laop$ and $\propset$ according to \Cref{def:kindop}, 
the principle of $\lak$-induction (cf.\ \Cref{cor:k_induction})
then tells us that%
\begin{align*}
    \lapp{\kindopfold{\propset}{\lak}(\propset)} ~{}\subseteq{}~ \propset 
    \qqimplies
    \underbrace{\lfp \laop}_{\mathclap{\text{reachable states of $\transsys$}}} ~{}\subseteq{}~ \propset~.
\end{align*}%
For our above choices, the premise of $\lak$-induction equals the classical formalization of $k$-induction---formulae \eqref{form:kind_init} and \eqref{form:kind_step}---because
the set of initial states $\init$ is \enquote{baked into} the operator $\laop$.
More concretely, for the base case \eqref{form:kind_init}, we have
\begin{align*}
	&\underbrace{
		\underbrace{
			\underbrace{
				\underbrace{
					\initf(s_1)
				}_{
					\lapp{\emptyset}
				} 
				\wedge 
				\transrelf(s_1,s_2)
			}_{
				\lappup{2}{\emptyset}
			} 
			\wedge 
			\ldots 
			\wedge 
			\transrelf(s_{k-1},s_k)
		}_{
			\lappup{k}{\emptyset}
		} 
		\implies \propsetf(s_1) \wedge \ldots \wedge \propsetf(s_k)
	}_{
		\qquad\qquad~~~~~\textnormal{meaning} \quad \lappup{k}{\emptyset} ~{}\subseteq{}~\propset
	}~.
	%
\end{align*}
    In other words, formula \eqref{form:kind_init} captures those states that are reachable from $\init$ via at most $k$ transitions.
    If we assume that \eqref{form:kind_init} is valid, then $\propset$ contains all initial states and formula \eqref{form:kind_step} coincides with the premise of $\lak$-induction:
\begin{align*}
	&
	\underbrace{
		\underbrace{
			\underbrace{
				\underbrace{
					\underbrace{
						\propsetf(s_1)\wedge \transrelf(s_1,s_2)
					}_{
						\lapp{\propset}
					} 
					\wedge 
					\propsetf(s_2)
				}_{
					\kindop{\propset}(\propset) \eeq \lapp{\propset} \cap \propset
				} 
				\wedge 
				\transrelf(s_2,s_3) 
				\wedge 
				\ldots 
				\wedge 
				\propset(s_k)
			}_{
				\kindopfold{\propset}{k-1}(\propset)
			} 
			\wedge 
			\transrelf(s_k,s_{k+1})
		}_{
			\lapp{\kindopfold{\propset}{k-1}(\propset)}
		}
		\implies 
		\propsetf(s_{k+1})
	}_{
		\qquad\qquad\qquad\qquad~\textnormal{meaning} \quad \lapp{\kindopfold{\propset}{k-1}(\propset)} ~{}\subseteq{}~ \propset
	}~.
\end{align*}%
It follows that, when considering transition systems, our (latticed) $\lak$-induction is equivalent to the classical notion of $k$-induction for $\lak < \omega$:
%
%
%
\begin{theorem}
\label{thm:latticed_vs_classical}
For every natural number $k \geq 1$,
\begin{align*}
	\lapp{\kindopfold{\propset}{k-1}(\propset)} ~{}\subseteq{}~ \propset 
	\qiff
    \textnormal{formulae \eqref{form:kind_init} and \eqref{form:kind_step} are valid}~. 
\end{align*}
\end{theorem}
\begin{proof}
	See Appendix~\ref{proof:thm:latticed_vs_classical}.
	\qed
\end{proof}

\section{Latticed Bounded Model Checking}
\label{sec:bmc}
%
%
We complement $\lak$-induction with a latticed analog of bounded model checking~\cite{DBLP:conf/tacas/BiereCCZ99,DBLP:series/faia/Biere09}
for \emph{refuting}
that $\lfp \laop \laord \laf$.
In lattice-theoretic terms, bounded model checking amounts to a \emph{fixed point iteration} of $\laop$ on $\bot$ while continually checking whether the iteration exceeds our candidate upper bound $\laf$.
If so, then we have indeed refuted $\lfp \laop \laord \laf$:%
\begin{theorem}[Soundness of Latticed BMC]
	Let $\laf \in \laset$. Then%
	\begin{align*}
        \exists\, \textnormal{ordinal } \od\colon \quad 
        \lappup{\od}{\bot} ~{}\not\laord{}~ \laf 
        \qqimplies 
        \lfp \laop~{}\not\laord{}~ \laf~.
	\end{align*}
\end{theorem}%
Furthermore, if we were actually able to perform transfinite iterations of $\laop$ on~$\bot$, 
then latticed bounded model checking is also complete:
If $\laf$ is in fact \emph{not} an upper bound on $\lfp \laop$, this \emph{will} be witnessed at some ordinal:%
\begin{theorem}[Completeness of Latticed BMC]
	Let $\laf \in \laset$. Then%
	\begin{align*}
		\lfp \laop ~{}\not\laord{}~ \laf 
		\qqimplies
        \exists\, \textnormal{ordinal } \od\colon\quad \lappup{\od}{\bot} ~{}\not\laord{}~ \laf~.
	\end{align*}%
\end{theorem}%
More practically relevant, if $\Phi$ is continuous (which is the case for Bellman operators characterizing reachability probabilities in Markov chains), then a simple \emph{finite} fixed point iteration, see Algorithm~\ref{alg:bmc}, is sound and complete for refutation:%
\begin{corollary}[Latticed BMC for Continuous Operators]
    \label{thm:latticed-bmc-continuous}
	Let $\laf \in \laset$ and let $\laop$ be continuous. Then%
	\begin{align*}
        \exists\, n \in \Nats\colon \quad
        \laop^{n}(\bot) ~{}\not\laord{}~ \laf
		\qqiff 
		\lfp \laop~{}\not\laord{}~\laf~.
	\end{align*}%
	%
	%
\end{corollary}

\section{Probabilistic Programs}
\label{sec:ppwp}


%

\begin{figure}[t]
\begin{adjustbox}{max width=\textwidth}
\begin{minipage}[b]{0.33\linewidth}
\begin{align*}
     \cc ~::=~ 
	& \SKIP	\\
    \qmid & \ASSIGN{x}{\ee} \\
    \qmid & \COMPOSE{\cc}{\cc} \\
    \qmid & \PCHOICE{\cc}{p}{\cc}  \\
	%
    \qmid & \ITE{\bb}{\cc}{\cc} \\
    \qmid & \WHILEDO{\bb}{\cc}
\end{align*}
\subcaption{$\pgcl$ programs}
\label{fig:grammars:pgcl}
\end{minipage}
\begin{minipage}[b]{0.43\linewidth}
\begin{align*}
\ee ~::=~
   & n \\ 
   \qmid & x \\ 
   \qmid & n \cdot \ee  \\ 
   \qmid & \ee+\ee  \\ 
    \qmid & \ee \monus \ee 
    \tag{monus $\max \{0, \ee - \ee\}$}
   \\
\end{align*}
\subcaption{Linear expressions}
\label{fig:grammars:expressions}
\end{minipage}
\begin{minipage}[b]{0.2\linewidth}
\begin{align*}
\bb ~::=~
   & \ee < \ee  \\ 
   \qmid & \bb \wedge \bb \\
   \qmid & \neg \bb 
    \\
    \\
    \\
\end{align*}
\subcaption{Linear guards}
\label{fig:grammars:guards}
\end{minipage}
\end{adjustbox}
\caption{
  Syntax of $\pgcl$ programs, linear expressions, and guards,
  where $x$ is a variable taken from a countable set $\Vars$ of program variables (evaluating to natural numbers),
  $p \in [0,1] \cap \Rats$ is a rational probability, 
  and $n \in \Nats$ is a constant.
}
\label{fig:grammars}
\end{figure}

%
%
In the remainder of this article, 
we employ latticed $k$-induction and BMC to verify imperative programs 
with access to discrete probabilistic choices---branching on the outcomes of coin flips.
In this section, 
we briefly recap the necessary background on formal reasoning about \mbox{probabilistic programs (cf.~\cite{Kozen1985,McIverM05} for details).}

\subsection{The Probabilistic Guarded Command Language}
%
\paragraph{Syntax.}
Programs in the \emph{probabilistic guarded command language} $\pgcl$ adhere to the grammar in \Cref{fig:grammars:pgcl}.
The semantics of most statements is standard.
In particular, the \emph{probabilistic choice} $\PCHOICE{\cc_1}{p}{\cc_2}$ flips a coin with bias $p \in [0,1] \cap \Rats$. If the coin yields heads, it executes $\cc_1$; otherwise, $\cc_2$.
%
In addition to the syntax in \Cref{fig:grammars}, we admit standard expressions that are definable as syntactic sugar, e.g., $\true$, $\false$, $\bb_1 \vee \bb_2$, $\ee_1 = \ee_2$, $\ee_1 \leq \ee_2$, etc.
%

%
\paragraph{Program states.}
A \emph{program state} $\pstate$ maps every variable in $\Vars$ to its value, i.e., a natural number in $\Nats$.\footnote{We prefer unsigned integers because our quantitative \enquote{specifications} 
(aka \emph{expectations}) must evaluate to non-negative numbers. Otherwise, expectations like $x+y$ are not well-defined, and, as a remedy, we would frequently have to take the absolute value of every program variable. Restricting ourselves to unsigned variables does not decrease expressive power as signed variables can be emulated (cf.~\cite[Sec.~11.2]{relatively_complete_prob_progs}).} 
To ensure that the set of program states $\States$ remains 
countable\footnote{In order to avoid any technical issues pertaining to measurability.}, we restrict ourselves to states in which only finitely many variables---those that appear in a given program---evaluate to 
non-zero values. Formally,
\[
    \States \ddefeq \setcomp{
        \pstate \colon \Vars \to \Nats
    }{ 
        \vphantom{\Bigl(} \bigl| \setcomp{ x \in \Vars }{ \pstate(x) \neq 0 } \bigr| < \infty
    }~.
\]
The evaluation of expressions $\ee$ and guards $\bb$ under a state $\pstate$, denoted by $\ee(\pstate)$ and $\bb(\pstate)$, is standard.
For example, we define the evaluation of ``monus'' as%
\belowdisplayskip=0pt%
 \begin{align*}
    (\ee_1 \monus \ee_2)(\pstate) \ddefeq 
    \max \left\{\,
        0,~ \ee_1(\pstate) - \ee_2(\pstate)
    \,\right\}~.
 \end{align*}%
\normalsize%
%


\subsection{Weakest Preexpectations}
\label{sec:wp}

\paragraph{Expectations.}
An \emph{expectation} $\ff \colon \States \to \PosRealsInf$ is a map from program states to the non-negative reals extended by infinity. 
We denote by $\E$ the set of all expectations.
Moreover, $(\E,\, \preceq)$ forms a complete lattice, where the partial order $\preceq$ is given by the pointwise application of the canonical ordering $\leq$ on $\PosRealsInf$, i.e.,
\[
\ff \ppreceq \fg \qqiff \forall\, \sigma  \in \States \colon\quad \ff(\sigma) \lleq \fg(\sigma)~.
\]
To conveniently describe expectations evaluating to some $\rr \in \PosRealsInf$ for every state, we slightly abuse notation and denote by $\rr$ the constant expectation $\lambda \pstate \mydot \rr$.
Similarly, given an arithmetic expression $\ee$, we denote by $\ee$ the expectation $\lambda \pstate \mydot \ee(\pstate)$.
The least element of $(\E,\, \preceq)$ is $0$ and the greatest element is $\infty$.
We employ the \emph{Iverson bracket} notation~\cite{Iverson1962} to cast Boolean expressions into expectations, i.e.,\
\begin{align*}
	\label{eqn:sem_iverson}
	\iverson{\bb} \eeq \lambda \sigma \mydot 
	\begin{cases}
		1 &~\text{if}~\bb(\sigma) = \true ~,\\
		0 &~\text{if}~\bb(\sigma)= \false~.
	\end{cases}
\end{align*}%
%
%
%
The \emph{weakest preexpectation transformer} 
%
	$\wpsymbol \colon \pgcl \to (\E \to \E)$
%
is defined
in \textnormal{\Cref{table:wp}}, where
$\fg\subst{x}{\ee}$ denotes the substitution of variable $x$ by expression $\ee$, i.e.,
\begin{align*}
   \fg\subst{x}{\ee} \defeq \lambda \pstate\mydot \fg(\pstate\statesubst{x}{\ee(\pstate)}), ~~\text{where}~~\pstate\statesubst{x}{\ee(\pstate)} \defeq \lambda y\mydot
   \begin{cases}
   	    \ee(\pstate) &\text{if}~y=x~, \\
   	   \sigma(y) &\text{otherwise}~.
   	\end{cases}
\end{align*}
 We call $\wp{\cc}{\fg}$ the \emph{weakest preexpectation} of program $\cc$ w.r.t.\ postexpectation $\fg$.
 The weakest preexpectation $\wp{\cc}{\fg}$ is itself an expectation of type $\E$, which maps each initial state $\pstate$ to the expected value of $\fg$ after running $\cc$ on~$\pstate$. 
 More formally, if $\mu_\cc^\pstate$ is the distribution over final states obtained by executing~$\cc$ on initial state $\pstate$, then for any postexpectation $\fg$~\cite{Kozen1985},
\begin{align*}
   \wp{\cc}{\fg}(\pstate) 
   \eeq
   \sum\nolimits_{\tau \in \States} \mu_\cc^\pstate(\tau) \cdot \fg(\tau)~.
\end{align*}%
%
For a gentle introduction to weakest preexpectations, see~\cite[Chap.~2 and 4]{benni_diss}.
\begin{table}[t]

	\renewcommand{\arraystretch}{1.5}

\begin{center}
\begin{tabular}{@{\hspace{1em}}l@{\hspace{2em}}l}
	\toprule
	$\boldsymbol{\cc}$			& $\boldsymbol{\textbf{\textsf{wp}}\,\left \llbracket \cc\right\rrbracket  \left(\fg \right)}$ \\
	\midrule
	$\SKIP$					& $\fg$ 																					\\
	$\ASSIGN{x}{\ee}$			& $\fg\subst{x}{\ee}$ \\
	$\COMPOSE{\cc_1}{\cc_2}$		& $\wp{\cc_1}{\vphantom{\big(}\wp{\cc_2}{\fg}}$ \\
	$\PCHOICE{\cc_1}{\pp}{\cc_2}$		& $\pp \cdot \wp{\cc_1}{\fg} + (1- \pp) \cdot \wp{\cc_2}{\fg}$ \\
	$\ITE{\guard}{\cc_1}{\cc_2}$		& $\iverson{\guard} \cdot \wp{\cc_1}{\fg} + \iverson{\neg \guard} \cdot \wp{\cc_2}{\fg}$ \\
    $\WHILEDO{\guard}{\cc'}$		& $\lfp \fh\mydot \iverson{\neg \guard} \cdot \fg + \iverson{\guard} \cdot \wp{\cc'}{\fh}$ \\
	\bottomrule
\end{tabular}
\end{center}
\caption{Rules defining the weakest preexpectation transformer.}
\label{table:wp}
\end{table}


\section{BMC and $k$-Induction for Probabilistic Programs}
\label{sec:application2pps}

We now instantiate latticed $\lak$-induction and BMC (as developed in 
\Cref{sec:problem_statement,sec:k_induction,sec:latticed-vs-classical,sec:bmc})
to enable verification of loops written in $\pgcl$; 
we discuss practical aspects later in \Cref{sec:linear_expectations,sec:deciding_entailments,sec:computing_minima} and \Cref{sec:implementation}.
For the next two sections, we fix a loop 
\begin{align*}
    \cloop \qeq  \WHILEDO{\guard}{\cbody}~.
\end{align*}
For simplicity, we assume that the loop body $\cbody$ is loop-free (every probabilistic program can be rewritten as a single while loop with loop-free body~\cite{DBLP:conf/tase/RabehajaS09}).

Given an expectation $\fg \in \E$ and a candidate upper bound~$\ff \in \E$ on the expected value of $\fg$ after executing $\cloop$ (i.e.\ $\wp{\cloop}{\fg}$), we will apply latticed verification techniques to check whether $\ff$ indeed upper-bounds $\wp{\cloop}{\fg}$.

To this end, we denote by $\laop$ the \emph{characteristic functional}
of $\cloop$ and $\fg$, i.e., 
\[
    \laop\colon \quad \E \tto \E, \qquad \fh \mmapsto 
    \iverson{\neg \guard} \cdot \fg + \iverson{\guard} \cdot \wp{\cbody}{\fh}~,
\]
whose least fixed point defines $\wp{\cloop}{\fg}$ (cf. \Cref{table:wp}).
We remark that $\laop$ is a monotonic---and in fact even continuous---operator
over the complete lattice $(\E, \, \eleq)$
(cf.~\Cref{sec:wp}).
In this lattice, the meet 
is a pointwise minimum, i.e.,%
\begin{align*}
    \fh \lameet \fhp \eeq \fh \emin \fhp \ddefeq \lambda \pstate \mydot \min\set{ \fh(\pstate), \fhp(\pstate) }~.
\end{align*}%
By \Cref{def:kindop}, $\laop$ and $\fg$ then induce the (continuous) $\lak$-induction operator%
\begin{align*}
    \kindop{\ff}\colon \quad \E \tto \E,
    \qquad
    \fh \mmapsto \lapp{\fh} \emin \ff~.
\end{align*}%
With this setup, we obtain the following proof rule for reasoning about probabilistic loops as an immediate consequence of 
\Cref{cor:k_induction}: 
\begin{corollary}[$\boldsymbol{k}$-Induction for \textsf{pGCL}] 
	For every natural number $k \in \Nats$, 
    \label{cor:kind_pgcl}
    \[ 
      \lapp{\kindopfold{\ff}{k}(\ff)} \ppreceq \ff
      \qimplies
      \wp{\cloop}{\fg} \ppreceq \ff~.
    \]
\end{corollary}%
Analogously, refuting that $\ff$ upper-bounds the expected value of $\fg$ after execution of $\cloop$ via bounded model checking is an instance of 
\Cref{thm:latticed-bmc-continuous}:%
\begin{corollary}[Bounded Model Checking for \textsf{pGCL}] 
    \label{cor:bmc_pgcl}
    \[ 
      \exists\, n \in \Nats\colon \quad \laop^{n}(0) ~{}\not\preceq{}~ \ff
	  \qqiff
      \wp{\cloop}{\fg} ~{}\not\preceq{}~ \ff~.
    \]
\end{corollary}%
%
%
%
\begin{example}[Geometric Loop]
\label{ex:pgcl_kind_bmc}
	The $\pgcl$ program%
	\begin{align*}
		\cgeo \qeq   \WHILEDO{x=1}{\PCHOICE{\ASSIGN{x}{0}}{0.5}{\ASSIGN{c}{c+1}}}
	\end{align*}%
	keeps flipping a fair coin $x$ until it flips 
    heads, sets $x$ to $0$, and terminates.
	Whenever it flips tails instead, it increments the counter $c$ and continues.
	We refer to $\cgeo$ as the \enquote{geometric loop} because after its execution, the counter variable $c$ is distributed according to a geometric distribution.
	
    What is a (preferably small) upper bound on the expected value
    $\wp{\cgeo}{c}$ of $c$ after execution of $\cgeo$?
    Using $2$-induction, we can (automatically) verify that $c+1$ is indeed an upper bound: 
    Since
    $\lapp{\kindop{c+1}(c+1)} \preceq c+1$,
    where $\laop$ denotes the characteristic functional of $\cgeo$,
    \Cref{cor:kind_pgcl} yields 
    $\wp{\cgeo}{c} \preceq c + 1$.

    However, $c+1$ \emph{cannot} be proven an upper bound using Park induction as it is \emph{not} 
    inductive.
    Moreover, it is indeed the \emph{least} upper bound, i.e., any smaller bound is refutable using BMC (cf.~\Cref{cor:bmc_pgcl}).
    For example, we have
    $\wp{\cgeo}{c} \not\preceq c + 0.99$, since
    $\lappup{11}{0} \not\preceq c+0.99$.
    Finally, we remark that some correct upper bounds only become $\lak$-inductive for \emph{transfinite} ordinals $\lak$. For instance, the innocuous-looking bound $2 \cdot c + 1$ 
    is not $k$-inductive for any natural number $k$, but it is $(\omega+1)$-inductive, since 
     $\laop \bigl(\kindopfold{2\cdot c+1}{\omega}(2 \cdot c+1) \bigr) \preceq 2\cdot c+1$. 
     \exped
\end{example}%
In principle, we can semi-decide whether $\wp{\cloop}{\fg} \not\preceq \ff$ holds or whether $\ff$ is $k$-inductive for some $k$: it suffices to run 
Algorithms~\ref{alg:k_induction} and~\ref{alg:bmc} in parallel.
However, for these two algorithms to actually be semi-decision procedures,
we cannot admit arbitrary expectations.
Rather, we restrict ourselves to a suitable subset $\SyntE$ of expectations in $\E$ satisfying all of the following requirements: 
%
\begin{enumerate}
	\item \label[requirement]{req:closedPhi}
		$\SyntE$ is closed under computing the characteristic functional $\laop$, i.e.,%
		\begin{align*}
			\forall\, \fh \in \SyntE\colon \quad \lapp{\fh}~\text{is computable and belongs to $\SyntE$} ~.
		\end{align*}%

	\item \label[requirement]{req:entail}
		Quantitative entailments between expectations in $\SyntE$ are decidable, i.e.,%
		\begin{align*}
			\forall\, \fh, \fh' \in \SyntE\colon \quad \text{it is decidable whether}~\fh \ppreceq \fh'~.
		\end{align*}%

		\item \label[requirement]{req:closedMeet}
			(For $k$-induction) $\SyntE$ is closed under computing meets, i.e.,%
			\begin{align*}
				\forall\, \fh, \fh' \in \SyntE\colon \quad \fh \emin \fh'~\text{is computable and belongs to}~\SyntE~.
			\end{align*}%
\end{enumerate}
Below, we show that \emph{linear expectations} meet all of the above requirements.


\subsection{Linear Expectations}
\label{sec:linear_expectations}
%

Recall from \Cref{fig:grammars:expressions} that we assume all expressions appearing
in $\pgcl$ programs to be linear.
For our fragment of syntactic expectations, we 
consider \emph{extended} linear expressions $\eeinfty$
that (1) are defined over \emph{rationals} instead of natural numbers and (2)
admit $\infty$ as a constant (but not as a \emph{sub}expression).
Formally, the set of extended linear expressions is given by the following grammar:
\begin{align*}
\eeinfty ~::=~ \ee \mid \infty 
\quad
\qquad
\ee ~::=~ r \mid x \mid r \cdot \ee \mid \ee + \ee \mid \ee \monus \ee  
    \tag{$r \in \PosRats$} 
\end{align*}
Similarly, we admit extended linear expressions (without $\infty$) in  linear guards~$\bb$.\footnote{We do not admit $\infty$ in guards for convenience. In principle, all comparisons with $\infty$ in guards can be removed by a simple preprocessing step.}
With these adjustments to expressions and guards in mind, 
the set $\Elin$ of \emph{linear expectations} is defined by the grammar
\begin{align*}
\fh ~::=~ & \eeinfty 
\qmid \iverson{\bb} \cdot \fh 
\qmid \fh + \fh ~. 
\end{align*}
%
%
We write $\fh = \fhp$ if $\fh$ and $\fhp$ are \emph{\underline{s}y\underline{ntacticall}y identical}; and $\fh \equiv \fhp$ if they are \emph{\underline{semanticall}y equivalent}, i.e.,
if for all states $\pstate$, we have $\fh(\pstate) = \fhp(\pstate)$.
Furthermore, the \emph{rescaling} $c \cdot \fh$ of a linear expectation $\fh$ by a constant $c \in \PosRats$ is syntactic sugar for 
rescaling suitable\footnote{We do not rescale every subexpression to account for the corner cases $c \cdot \infty = \infty$ and $0 \cdot \infty = 0$.}
arithmetic subexpressions of $\fh$, e.g.,
\[ 
{\color{blue}\sfrac{1}{2}} \cdot (\iverson{x = 1} \cdot 4 + \sfrac{1}{3} \cdot x + \infty) 
~\equiv~
{\color{blue}\sfrac{1}{2}} \cdot \iverson{x = 1} \cdot 4 + {\color{blue}\sfrac{1}{2}} \cdot \sfrac{1}{3} \cdot x + \infty \in \Elin~.
\]
A formal definition of the rescaling $c \cdot \fh$ is found in \Cref{app:linear-rescaling}.

If we choose a linear expectation $\fh$ as a postexpectation, then a quick inspection of \Cref{table:wp} reveals that the weakest preexpectation $\wp{\cc}{\fh}$ of any \emph{loop-free} $\pgcl$ program $\cc$ and $\fh$ yields a linear expectation again.
Hence, linear expectations are closed under applying $\laop$---\Cref{req:closedPhi} above---because
\begin{align*}
    \forall\, \fg,\fh \in \Elin\colon\quad
    \lapp{\fh} \eeq 
    \underbrace{
      \underbrace{\iverson{\neg \guard} \cdot \fg}_{~\in~ \Elin} 
      \,+\, \iverson{\guard} \cdot \underbrace{\wp{\cbody}{\fh}}_{~\in~ \Elin}
    }_{~\in~ \Elin}~.
\end{align*}

\subsection{Deciding Quantitative Entailments between Linear Expectations}
\label{sec:deciding_entailments}

To prove that linear expectations meet \Cref{req:entail}---decidability of quantitative entailments---we effectively reduce the question of whether an entailment $\fh \preceq \fhp$ holds to the decidable satisfiability problem for \qflira---quantifier-free mixed linear integer and real arithmetic (cf.~\cite{DBLP:conf/smt/0001BT14}).

As a first step, we show that every linear expectation can be represented as a sum of mutually exclusive extended arithmetic expressions---a representation we refer to as the \emph{guarded normal form} (similar to~\cite[Lemma $1$]{DBLP:conf/sas/KatoenMMM10}).
\begin{definition}[Guarded Normal Form (GNF)]
	\label{def:gnf}
    $\fh \in \Elin$ is in GNF if 
    \begin{align*}
        \fh \eeq \sum\nolimits_{i = 1}^{n} \iverson{\bb_i} \cdot \eeinfty_i~,
    \end{align*}
    where
    $\eeinfty_1,\ldots,\eeinfty_n$ are extended linear expressions, 
    $n \in \Nats$ is some natural number,
    and
    $\bb_1, \ldots, \bb_n$ are linear Boolean expressions that partition the set of states, i.e., for each $\pstate \in \States$ there exists \emph{exactly one} $i \in \{1,\ldots,n\}$ such that $\bb_i (\pstate) = \true$.
\end{definition}%
\begin{lemma}
	\label{lem:gnf}
  Every linear expectation $\fh \in \Elin$ can effectively be transformed 
    into an equivalent linear expectation $\gnf{\fh} \equiv \fh$ in guarded normal form.
\end{lemma}%
\begin{proof} An extension of \cite[Lemma A.2]{relatively_complete_prob_progs}. See details in \Cref{app:gnf}. \end{proof}
The number of summands $|\gnf{\fh}|$ in $\gnf{\fh}$ is, in general, exponential in the number of summands in $\fh$. 
In practice, however, this exponential blow-up can often be mitigated by pruning summands with unsatisfiable guards.
%
Throughout the remainder of this paper, we denote the components of
$\gnf{\fh}$ and $\gnf{\fhp}$, where $\fh$ and $\fhp$ are arbitrary linear expectations, 
as follows:
\begin{align*}
	\gnf{\fh} \eeq \sum\nolimits_{i=1}^{n} \iverson{\bb_i}\cdot \eeinfty_i 
	\qand
	\gnf{\fhp} \eeq& \sum\nolimits_{j=1}^{m} \iverson{\bp_j}\cdot \eainfty_j ~.
\end{align*}
We now present a decision procedure for the \emph{quantitative entailment} over $\Elin$.%
\begin{theorem}[Decidability of Quantitative Entailment over \textsf{LinExp}]
\label{thm:entailment_elin}%
For $\fh,\fhp \in \Elin$, it is decidable whether $\fh \preceq \fhp$ holds.
\end{theorem}%
\begin{proof}
Let $\fh,\fhp \in \Elin$.
By \Cref{lem:gnf}, we have $\fh \preceq \fhp$ iff $\gnf{\fh} \preceq \gnf{\fhp}$.

Let $\pstate$ be some state.
By definition of the GNF, $\pstate$ satisfies
exactly one guard $\bb_i$ and exactly one guard $\bp_j$.
Hence, the inequality $\gnf{\fh}(\sigma) \leq \gnf{\fhp}(\sigma)$ does \emph{not} hold
iff $\eeinfty_i(\sigma) > \eainfty_j(\sigma)$ holds for the expressions $\eeinfty_i$ and $\eainfty_j$ guarded by $\bb_i$ and $\bp_j$, respectively.
Based on this observation, we construct a $\qflira$ formula
$\smtentails{\fh}{\fhp}$
that is \emph{unsatisfiable} iff there is no counterexample to the entailment $\fh \preceq \fhp$
(see Appendix~\ref{app:entail-proof} for a soundness proof):
%
\begin{align*}
	\smtentails{\fh}{\fhp} 
	 \ddefeq &
	\bigvee\nolimits_{i=1}^n ~
		\bigvee\nolimits_{\substack{j=1,\, \eainfty_j \neq \infty}}^m 
			\left(\bb_i \wwedge\bp_j 
					\wwedge               \tosmt{\eeinfty_i}       >  \eainfty_j
			\right) ~.
\end{align*}
Here, we identify every program variable in $\fh$ or $\fhp$ with an $\Nats$-valued SMT variable.
Moreover, to account for comparisons with $\infty$, we rely on the fact that our (extended) arithmetic expressions either evaluate to $\infty$ for \emph{every} state or \emph{never} evaluate to $\infty$.
To deal with the case $\eeinfty_i > \infty $, which is always false, 
we can thus safely exclude cases in which $\eainfty_j = \infty$ holds.
To deal with the case $\infty > \eainfty_j$, we represent $\infty$ by some unbounded number, i.e., we introduce a fresh, unconstrained $\Nats$-valued SMT 
variable $\smtinfty$ and set
$\tosmt{\eeinfty}$ to $\smtinfty$ if $\eeinfty = \infty$; otherwise, $\tosmt{\eeinfty} = \eeinfty$.
Since $\qflira$ is decidable (cf.~\cite{DBLP:conf/smt/0001BT14}), we conclude that
the quantitative entailment problem is decidable.
\qed
\end{proof}%
Since quantitative entailments are decidable, we can already conclude that, for linear expectations,
\Cref{alg:bmc} is a semi-decision procedure.%
%
%
%

\subsection{Computing Minima of Linear Expectations}
\label{sec:computing_minima}

To ensure that latticed $k$-induction on $\pgcl$ programs (cf.~\Cref{alg:k_induction} and~\Cref{sec:application2pps}) is a semi-decision procedure when considering linear expectations, we have to consider \Cref{req:closedMeet}---the expressability and computability of meets:%
\begin{theorem}
	\label{thm:min_closed_elin}
	$\Elin$ is effectively closed under taking minima.
\end{theorem}%
%
%
\begin{proof}
    For $k\in \Nats$, let $\onetoarg{k} \defeq \{1,\ldots,k\}$. 
    Then, for two linear expectations $\fh, \fhp$, the
    linear expectation $\gnf{\fh} \emin \gnf{\fhp} \in \Elin$ is given by: 
	%
   %
   %
   \begin{align*}
   	   \sum_{(i,j)\, \in\, \onetoarg{n}\times\onetoarg{m}}
   	   \begin{cases}
   	   \iverson{\bb_i \wedge \bp_j} \cdot \eainfty_j, &~\text{if $\eeinfty_i = \infty$}~, \\
   	   \iverson{\bb_i \wedge \bp_j} \cdot \eeinfty_i, &~\text{if $\eainfty_i = \infty$}~, \\
   	    \iverson{\bb_i \wedge \bp_j \wedge \eeinfty_i \leq \eainfty_j} \cdot \eeinfty_i
   	    + 
   	    \iverson{\bb_i \wedge \bp_j \wedge \eeinfty_i  > \eainfty_j} \cdot \eainfty_j &~\text{otherwise}~,
   	   	\end{cases}
   \end{align*}
    where we exploit that, for every state, exactly one guard $\bb_i$ and exactly one guard $\bp_j$ is satisfied (cf. \Cref{lem:gnf}).
    Notice that in the last case we indeed obtain a linear expectation since neither $\eeinfty$ nor $\eainfty$ are equal to $\infty$.
    \qed
\end{proof}%
%
%
In summary, all requirements stated in \Cref{sec:application2pps} are satisfied.
%
%
%
%
%
%

%

\section{Implementation}
\label{sec:implementation}

\newcommand{\unintfun}{P}
\newcommand{\unintfunk}{Q}
\newcommand{\encoding}[1]{\rho_{#1}}

We have implemented a prototype called $\tool$---$k$-Induction for PRObabilistic PRO\-grams---in Python $3.7$ using the SMT solver Z3~\cite{z3} and the solver-API PySMT~\cite{pysmt2015}. 
Our tool, its source code, and our experiments are available online.\footnote{\faGithub~\url{https://github.com/moves-rwth/kipro2}}
$\tool$ performs in parallel latticed $k$-induction and BMC to fully automatically verify upper bounds on expected values of $\pgcl$ programs as described in \Cref{sec:application2pps}.
In addition to reasoning about expected values, $\tool$ supports verifying bounds on \emph{expected runtimes} of $\pgcl$ programs, which are characterized as least fixed points \`a la \cite{ert_journal}. 
Rather than fixing a specific runtime model, we took inspiration from~\cite{DBLP:conf/pldi/NgoC018} and added a statement $\TICK{n}$
that does not affect the program state but consumes $n \in \Nats$ time units.

To discharge quantitative entailments and compute the meet, we use the constructions in \Cref{thm:entailment_elin,thm:min_closed_elin}, respectively.
As an additional optimization, we do not iteratively apply the $k$-induction operator $\kindop{\laf}$ directly but use an \emph{incremental encoding}.
%
%
We briefly sketch 
our encoding
for
$k$-induction (Algorithm~\ref{alg:bmc}); the encoding for BMC is similar. 
In both cases, we employ uninterpreted functions on top of mixed integer and real arithmetic, i.e., 
$\qfuflira$.

Recall \Cref{ex:pgcl_kind_bmc}, the geometric loop $\cgeo$, where we used $k$-induction to prove 
$\wp{\cgeo}{c} \preceq c + 1$.
For every $k \in \Nats$, 
$\laop(\kindopfold{c+1}{k}(c+1))$ is given by
\begin{align*}
	%
	%
	\underbrace{\iverson{x=1}\cdot \Big(0.5 \cdot \underbrace{\kindopfold{c+1}{k}(c+1)}_{\unintfunk_k}\subst{x}{0} \,+\, 0.5 \cdot \underbrace{\kindopfold{c+1}{k}(c+1)}_{\unintfunk_k}\subst{c}{c+1}\Big) \,+\, \iverson{x \neq 1}\cdot c}_{\unintfun_{k}}~. \notag
\end{align*}
%
To obtain an incremental encoding, we introduce an uninterpreted function
$\unintfun_k\colon \Nats \times \Nats \to \PosReals$
and a formula $\encoding{k}(c,x)$ specifying that $\unintfun_k(c,x)$
characterizes $\laop(\kindopfold{c+1}{k}(c+1))$, i.e., 
for all $\pstate \in \States$ and $r \in \PosReals$
with $\laop(\kindopfold{c+1}{k}(c+1))(\sigma)  < \infty$,\footnote{%
Notice that we do \emph{not} axiomatize in $\encoding{k}(c,x)$ that $\laop(\kindopfold{c+1}{k}(c+1))$
and $\unintfun_k(c,x)$ are the same function because we have no access to universal quantifiers.
Rather, we specify that both functions coincide for any fixed concrete values assigned to~$c$~and~$x$.
This weaker notion is \emph{not} robust against formal modifications of the parameters, e.g., through substitution.
For example, to assign the correct interpretation to $\unintfun_k(c,x)\subst{c}{c+1}$, we have to construct a (second) formula $\encoding{k}(c,x)\subst{c}{c+1}$.
}%
\begin{align*}
    &\encoding{k}(\sigma(c), \sigma(x)) \wedge \unintfun_k(\sigma(c), \sigma(x)) = r ~ \text{is satisfiable}  
	\qiff
	r=\laop\Big(\kindopfold{c+1}{k}(c+1)\Big)(\sigma)~.
\end{align*}%
If $\laop(\kindopfold{c+1}{k}(c+1))(\sigma)  = \infty$,
our construction of $\encoding{k}(x,c)$ ensures that 
the above conjunction is satisfiable for arbitrarily large $r$.
Analogously, we introduce an uninterpreted function $\unintfunk_k\colon \Nats \times \Nats \to \PosReals$ that characterizes $\kindopfold{c+1}{k}(c+1)$.
In particular, the formula $\encoding{k}(c,x)$ may
use all uninterpreted functions introduced for smaller or equal values of $k$---not just the function $\unintfun_k(c,x)$ it needs to characterize.
This enables an incremental encoding in the sense that
$\encoding{k}(c,x)$ can be computed on top of $\encoding{k-1}(c,x)$ 
by reusing $\unintfun_{k-1}(c,x)$, $\unintfunk_k(c,x)$, 
and the construction in \Cref{thm:min_closed_elin}.

Moreover, we can reuse $\encoding{k}(c,x)$ 
to avoid computing the (expensive) GNF for deciding certain quantitative entailments 
(cf.~Theorem~\ref{thm:entailment_elin}):
For example, to check whether
$\laop(\kindopfold{c+1}{k}(c+1)) \not\preceq \fhp$ holds, 
we only need to transform the right-hand side into GNF (cf.~\Cref{sec:deciding_entailments}),
i.e., if $\gnf{\fhp} = \sum_{j=1}^{m} \iverson{\psi_j} \cdot \eainfty_j$, then
\begin{align*}
    \laop\Big(\kindopfold{c+1}{k}(c+1)\Big) \not\preceq \fg
    \qiff
    \encoding{k} \wedge \bigvee\nolimits_{\substack{j=1, \, \eainfty_j \neq \infty}}^m  \psi_j \wedge \unintfun_k(c,x) > \eainfty_j~\text{is satisfiable}~.
\end{align*}

\section{Experiments}
\label{sec:impl:experiments}

We evaluate $\tool$ on two sets of benchmarks.
The first set, shown in \Cref{tab:results_wp}, consists of four (infinite-state) probabilistic systems compiled from the literature; each benchmark is evaluated on multiple variants of candidate upper bounds:

(1) \expname{brp} is a $\pgcl$ variant of the bounded retransmission protocol~\cite{DBLP:conf/types/HelminkSV93,DBLP:conf/papm/DArgenioJJL01}. 
The goal is to transmit $\vartosend$ many packages via an unreliable channel allowing for at most $\varmaxfailed$ many retransmissions per package (cf.~Example~\ref{ex:brp}). 
The variable $\vartotalfailed$ keeps track of the total number of failed attempts to send a package. 
We verified upper bounds on the expected outcome of $\vartotalfailed$ (variants $1$--$4$). 
In doing so, we bound the number of packages to send by $4$ ($10$, $20$, $70$) while keeping $\varmaxfailed$ \emph{unbounded}, i.e., we still verify an infinite-state system.
We notice that $k > 1$ is required for proving any of the candidate bounds; 
for up to $k = 11$, $\tool$ manages to prove non-trivial bounds within a few seconds.
However, unsurprisingly, the complexity increases rapidly with larger~$k$.
While $\tool$ can prove variant 3, it needs to increase $k$ to 23; we observe that the complexity grows rapidly both in terms of the size of formulae and in terms of runtime with increased $k$.
Furthermore, variants $5$--$7$ correspond to (increasing) incorrect candidate bounds ($\vartotalfailed+1$, $\vartotalfailed+1.5$, $\vartotalfailed+3$) that are refuted (or time out)
when not imposing any restriction on $\vartosend$.


(2) \expname{geo} corresponds to the geometric loop from \Cref{ex:pgcl_kind_bmc}.
We verify that $c+1$ upper-bounds the expected value of $c$ for every initial state (variant $1$);
we refute the incorrect candidates $c+0.99$ and $c+0.999999999999$ (variants $2$--$3$).

(3) \expname{rabin} is a variant of Rabin's mutual exclusion algorithm~\cite{DBLP:conf/podc/KushilevitzR92} taken from~\cite{DBLP:journals/tcs/HurdMM05}.
We aim to verify that the probability of obtaining a unique winning process is at most $\nicefrac{2}{3}$
for at most $2$ ($3$, $4$) participants (variants $1$--$3$) and refute both $\nicefrac{1}{3}$ (variant $4$) and \nicefrac{3}{5} (variant $5$) for an unbounded number of participants.

(4) \expname{unif\kern-.7pt\_gen} implements the algorithm in~\cite{DBLP:journals/corr/abs-1304-1916} for
generating a discrete uniform distribution over some interval $\{ \varfont{l}, \ldots,\varfont{l}+\varfont{n}-1 \}$ using only fair coin flips.
We aim to verify that $\nicefrac{1}{n}$ upper-bounds the probability of sampling a particular element from \emph{any} such interval of size at most $n=2$ ($3$, $4$, $5$, $6$) (variants $1$--$5$).

Our second set of benchmarks, shown in \Cref{tab:results_ert}, confirms the correctness of (1-inductive) bounds on the expected runtime of $\pgcl$ programs synthesized by the runtime analyzers \textsc{Absynth}~\cite{DBLP:conf/pldi/NgoC018} and (later) \textsc{KoAT}~\cite{koat}; this gives a baseline for evaluating the performance of our implementation. 
Moreover, it demonstrates the flexibility of our approach as we effortlessly apply the expected runtime calculus~\cite{ert_journal} instead of the weakest preexpectation calculus for verification.

Further details about individual benchmarks, including all considered $\pgcl$ programs and candidate upper bounds, can be found in \Cref{app:benchmarks}.

\paragraph{Setup.}
We ran Algorithms~\ref{alg:k_induction} and \ref{alg:bmc} in parallel using an AMD Ryzen 5 3600X processor with a shared memory limit of $8$GB and a 15-minute timeout. 
For every benchmark finishing within the time limit, $\tool$ either finds the smallest~$k$ required to prove the candidate bound by $k$-induction or the smallest unrolling depth~$k$ to refute it. 
If $\tool$ refutes, the SMT solver provides a concrete initial state witnessing that violation. 
In \Cref{tab:results_wp,tab:results_ert},
column \textsf{\#formulae} gives the maximal number of conjuncts on the solver stack; \textsf{formulae\_t}, \textsf{sat\_t}, and \textsf{total\_t} give the amount of time spent on (1) computing formulae, (2) satisfiability checking, and (3)~everything (including preprocessing), respectively. 
The input consists of a program, a candidate upper bound, and a postexpectation; in \Cref{tab:results_ert}, the latter is fixed to \enquote{postruntime} $0$ and thus omitted.

\begin{table}[t]
	\caption{Empirical results for the first benchmark set (time in seconds).}
		\label{tab:results_wp}
\centering
\adjustbox{width=0.85\textwidth}{
\begin{tabular}{lccrrrrrr}
	%
	\toprule
    &                                                ~postexpectation~ &  ~variant~ &                   ~~result &  ~~~~$k$ & ~~\textsf{\#formulae} & ~~\textsf{formulae\_t} &  ~~\textsf{sat\_t} &  ~~\textsf{total\_t} \\
	\midrule
	 \multirow{7}{*}{\rotatebox{90}{\expname{brp}}} & \multirow{7}{*}{$\vartotalfailed$}  & $1$   &ind &  5&        285 &                   0.15 &            0.01 &        0.28 \\
	 &  & $2$   &  ind & 11  &       2812 &                   1.77 &            0.12 &        2.03 \\
	& & $3$  &  ind & 23 &      26284 &                  17.68 &           28.09 &       45.94 \\
	 & & $4$ & TO &  -- &    --&                  -- &               -- &           -- \\
    &    &                                $5$ &   ref & 13 &        949 &                   0.84 &           14.39 &       15.28 \\
	 &    &                              $6$  &   TO &  -- &                      -- &               -- &           -- & --  \\
	&     &                               $7$ &   TO &  -- &                      -- &               -- &           --  & -- \\
	\midrule
	\multirow{3}{*}{\,\rotatebox{90}{\expname{geo}}} &     \multirow{3}{*}{$c$}                 &                          $1$  &  ind &  2 &18 &                   0.01 &            0.00 &        0.08 \\
	 &                     &                       $2$ &            ref & 11 &        103 &                   0.04 &            0.01 &        0.09 \\
    &                   &               $3$&              ref & 46 &       1223 &                   0.39 &            0.04 &        0.48 \\
	 \midrule
	\multirow{5}{*}{\rotatebox{90}{\expname{rabin}}}  & \multirow{5}{*}{$[i=1]$} & $1$ &        ind &  1  &         21 &                   0.01 &            0.00 &        0.15 \\
	 && $2$ &       ind &  5  &       1796 &                   1.27 &            0.03 &        1.44 \\
	 && $3$ &       TO &  -- &      --&                -- &               -- &           -- \\
&& $4$ &      ref &  4 &        458 &                   0.31 &            0.03 &        0.40 \\
	&& $5$&        ref &  8  &      10508 &                   8.76 &            2.85 &       11.68 \\
	\midrule
	\multirow{5}{*}{\rotatebox{90}{\expname{unif\kern-.7pt\_gen}}}& \multirow{5}{*}{$[c=i]$} &  $1$ &        ind &  2  &        267 &                   0.27 &            0.02 &        0.56 \\
	&& $2$ &         ind &  3  &       1402 &                   1.45 &            0.10 &        1.81 \\
	 && $3$&        ind &  3&       1402 &                   1.48 &            0.11 &        1.86 \\
	 && $4$ &      ind &  5  &      40568 &                  47.31 &           15.70 &       63.28 \\
	&& $5$ &        TO &  -- &                      -- &               -- &           -- & -- \\
	\bottomrule
\end{tabular}
}
\end{table}

\paragraph{Evaluation of Benchmark Set 1.}
%
Table~\ref{tab:results_wp} empirically underlines that probabilistic program verification can benefit from $k$-induction to the same extent as classical software verification: $\tool$ \emph{fully automatically} verifies relevant properties of \emph{infinite-state} randomized algorithms and stochastic processes from the literature that require \emph{$k$~to be strictly larger than $1$}. 
That is, proving these properties using ($1$-)inductive invariants 
requires either non-trivial invariant synthesis or additional user annotations. 
This indicates that $k$-induction mitigates the need for complicated specifications in probabilistic program verification (cf.~\cite{ert_journal}).

We observe that $k$-induction tends to succeed if \emph{some} variable is bounded 
in the candidate upper bound
under consideration (cf.\ \expname{brp}, \expname{rabin}, \expname{unif\kern-.7pt\_gen}). However, $k$-induction can also succeed without any bounds 
(cf.\ \expname{geo}). The time and formulae required for checking $k$-inductivity increases rapidly for larger $k$; this is particularly striking for \expname{rabin} and \expname{unif\kern-.7pt\_gen}.
When refuting candidate bounds with BMC, we obtain a similar picture. Both the time and formulae required for refutation increase if the candidate bound \mbox{increases (cf.\ \expname{brp}, \expname{geo}, \expname{rabin}).}

For both $k$-induction and BMC, we observe a direct correlation between the complexity of the loop, i.e., the number of possible traces through the loop from some fixed initial state after some bounded number of iterations, and the required time and space (number of formulae). Whereas for \expname{geo} and \expname{brp}---which exhibit a rather simple structure---these checks tend to be fast, this is not the case for \expname{rabin} and \expname{unif\kern-.7pt\_gen}, which have more complex loop bodies. For such complex loops, $k$-induction and BMC quickly become infeasible as $k$ increases.

\begin{table}[t]
	\caption{Empirical results for (a subset of) the ERTs~\cite{DBLP:conf/pldi/NgoC018} (time in \emph{milliseconds}).}
	\label{tab:results_ert}
\centering
\adjustbox{max width=\textwidth}{
\begin{tabular}{llr r r r r r}
	%
	\toprule
    &                       runtime bound candidate &             result & ~~$k$ &  ~\textsf{\#formulae} &  \textsf{formulae\_t} &  \textsf{sat\_t}&  \textsf{total\_t} \\
	\midrule
	\expname{2drwalk} &                 $2\cdot (n+1 \monus d)$ &  TO   & -- &             -- &                      -- &               -- &           -- \\
	\expname{bayesian\kern-.7pt\_network} &                       $5\cdot n$ &  TO  & -- &             -- &                      -- &               -- &           -- \\
	\expname{ber} &                   $2\cdot (n\monus x)$ & ind & 1 &          9 &                   7.22 &            0.44 &       88.12 \\
	\expname{C4B\kern-.7pt\_t303} &     $0.5\cdot (x+2) + 0.5\cdot (y+2)$ & ind & 3 &        129 &                 91.38 &           10.01 &      216.11 \\
	\expname{condand} &                       $m+n$ & ind & 1 &         10 &                     7.10 &            0.43 &       76.21 \\
	\expname{fcall} &                   $2\cdot (n\monus x)$ & ind & 1 &          9 &                6.73 &            0.41 &       75.73 \\
	\expname{hyper} &                   $5\cdot (n\monus x)$ & ind & 1 &         11 &                    7.24 &            0.46 &       97.52 \\
	\expname{linear01} &                     $0.6\cdot x$ & ind & 1 &         11 &                  7.19 &            0.49 &       74.38 \\
	\expname{prdwalk} &           $1.14286\cdot (n+4\monus x)$ & ind & 1 &         17 &                  7.64 &            0.72 &      194.44 \\
	\expname{prspeed} & $2\cdot (m\monus y) + 0.6666667\cdot (n\monus x)$ & ind & 1 &         18 &                    7.64 &            0.81 &      145.13 \\
	\expname{race} &         $ 0.666667\cdot (t+9\monus h)$ & ind & 1 &         30 &                     9.21 &            0.86 &      695.89 \\
	\expname{rdspeed} &    $2\cdot (m\monus y)+0.666667\cdot (n\monus x)$ & ind & 1 &         19 &                   7.70 &            0.78 &      143.45 \\
	\expname{rdwalk} &                $ 2\cdot (n+1\monus x)$ & ind & 1 &         12 &               10.22 &            0.75 &       85.03 \\
	\expname{sprdwalk} &                   $2\cdot (n\monus x)$ & ind & 1 &          9 &             7.28 &            0.42 &       83.40 \\
	\bottomrule
\end{tabular}

}
\end{table}

\paragraph{Evaluation of Benchmark Set 2.}
%
From \Cref{tab:results_ert}, we observe that---in almost every case---verification is instantaneous and requires very few formulae. 
The programs we verify are equivalent to the programs provided in~\cite{DBLP:conf/pldi/NgoC018} up to interpreting minus as \emph{monus} and using $\Nats$-typed (instead of $\Ints$) variables. A manual inspection reveals that this matters for \expname{C4B\kern-.7pt\_t303} and \expname{rdwalk}, which is the reason why the runtime bound for \expname{C4B\kern-.7pt\_t303} is $3$-inductive rather than $1$-inductive.

There are two timeouts (\expname{2drwalk}, \expname{bayesian\kern-.7pt\_network}) due to the GNF construction from Lemma~\ref{lem:gnf}, which exhibits a runtime exponential in the number of possible execution branches through the loop body. We conjecture that further preprocessing (by pruning infeasible branches upfront) can mitigate this, rendering \expname{2drwalk} and \expname{bayesian\kern-.7pt\_network} tractable as well. 
We consider a thorough investigation of suitable preprocessing strategies for GNF construction, which is outside the scope of this paper, a worthwhile direction for future research.

\section{Conclusion}
\label{sec:conclusion}
%
%
We presented $\lak$-induction, a generalization of classical $k$-induction to arbitrary complete lattices, and---together with a complementary bounded model checking approach---obtained a fully automated technique for verifying infinite-state probabilistic programs. 
Experiments showed that this technique can prove non-trivial properties in an automated manner that using existing techniques cannot be proven---at least not without synthesizing a stronger inductive invariant.
If a given candidate bound is $k$-inductive for some $k$, then our
prototypical tool will find that $k$ for linear programs and linear expectations.
In theory, our tool is also applicable to non-linear programs at the expense of 
an undecidability quantitative entailment problem.
It is left for future work to consider (positive) real-valued program variables for non-linear expectations.

\subsubsection*{Acknowledgements.}
Benjamin Lucien Kaminski is indebted to Larry~Fischer for his linguistic advice---this time on the word \enquote{latticed}.

\bibliographystyle{splncs04}

\bibliography{literature}


\appendix
\section{Appendix}

\subsection{Proof of \Cref{lem:kindop_mon_chain}}
\label{proof:lem:kindop_mon_chain}
	%
    %
	For \cref{lem:kindop_mon_chain_1}, observe that $\lah_1 \laord \lah_2$ implies $\lah_1 \sqcap \laf \laord \lah_2 \sqcap \laf$. We then have
	\begin{align*}
        \kindop{\laf}(\lag_1) \eeq& \lapp{\lag_1} \sqcap \laf
	   \tag{by definition of $\kindop{\laf}$} \\
       ~{}\laord{}~& \lapp{\lag_2} \sqcap \laf
	   \tag{by monotonicity of $\Phi$ and the above property} \\
	   \eeq& \kindop{\laf}(\lag_2)~.
	   \tag{by definition of $\kindop{\laf}$}
	\end{align*}
	For \cref{lem:kindop_mon_chain_2}, we proceed by transfinite induction on the $\oc$. \\ \\
	\emph{The case $\od=0$.} 
    This case is trivial as there exists no ordinal $\oc < \od$.
    \\
    \emph{The case $\od = \ob + 1$ successor ordinal.}
    For every $\oa < \ob$, consider the following:
	\begin{align*}
        & \kindopfold{\laf}{\od}(\laf) \\
        \eeq &
        \kindopapp{\laf}{
          \kindopfold{\laf}{\ob}(\laf)
        }
        \tag{by definition of $\kindopfold{\laf}{\ob+1}(\laf)$} \\
        \llaord &
        \kindopapp{\laf}{
          \kindopfold{\laf}{\oa}(\laf)
        }
        \tag{by I.H. and monotonicity of $\kindop{\laf}$} \\
        \eeq &
        \kindopfold{\laf}{\oa+1}(\laf)~.
        \tag{by definition of $\kindopfold{\laf}{\oa+1}(\laf)$}
	\end{align*}
    This proves the claim for every $\oc = \oa + 1 < \od$.
    For the missing case $\oc = 0$, consider the following:
	\begin{align*}
        & \kindopfold{\laf}{\od}(\laf) \\
        \eeq &
        \kindopapp{\laf}{
          \kindopfold{\laf}{\ob}(\laf)
        }
        \tag{by definition of $\kindopfold{\laf}{\ob+1}(\laf)$} \\
        \llaord & \laf 
        \tag{by \Cref{lem:kindop_mon_chain}, definition of $\kindop{\lag}$} \\
        %
        \eeq & \kindopfold{\laf}{0}(\laf)~.
        \tag{by definition of $\kindopfold{\laf}{0}(\laf)$}
	\end{align*}
    \emph{The case $\od$ limit ordinal.}
    For every $\oc < \od$ consider the following
	\begin{align*}
        & \kindopfold{\laf}{\od}(\laf) \\
        \eeq &
          \inf \left\{ \kindopfold{\laf}{\ob}(\laf) \mathrel{\big|} \ob < \od \right\}
        \tag{by definition of $\kindopfold{\laf}{\od}(\laf)$} \\
        \llaord &
          \inf \left\{ \kindopfold{\laf}{\oa}(\laf) \mathrel{\big|} \oa < \oc \right\}
        \tag{by I.H.} \\
        \eeq & \kindopfold{\laf}{\oc}(\laf)~.
        \tag{by definition of $\kindopfold{\laf}{\oc}(\laf)$} 
	\end{align*}
    For \cref{lem:kindop_mon_chain_4}, 
    we first observe that $\kindopapp{\laf}{\lag} \laord \lapp{\lag}$
    holds for every element $\lag \in \laset$.
    The claim then follows from a straightforward transfinite induction on
    the number of iterations and the fact that
    $\lappdown{\od}{\lag} \laord \lappup{\od}{\lag}$ holds by definition.
	For \cref{lem:kindop_mon_chain_3}, 
    assume $\lfp \laop \laord \laf$.
    It suffices to prove that, for all ordinals $\od < \lambient$, we have 
    $\lfp \laop \laord \kindopfold{\laf}{\od}(\laf)$; the remaining 
    inequalities are immediate by definition of $\kindop{\laf}$
    and \cref{lem:kindop_mon_chain_2}.
    We proceed by transfinite induction on $\od$.

    \emph{The case $\od = 0$.}
    Trivial, since 
    $\lfp \laop \llaord \laf \eeq \kindopfold{\laf}{0}(\laf)$.
    
    \emph{The case $\od = \oc+1$ successor ordinal.}
	\begin{align*}
        & \kindopfold{\laf}{\od}(\laf) \\
        \eeq &
        \kindopapp{\laf}{
          \kindopfold{\laf}{\oc}(\laf)
        }
        \tag{by definition of $\kindopfold{\laf}{\od}(\laf)$} \\
        \llaordrev &
        \kindopapp{\laf}{
          \lfp \laop
        }
        \tag{by I.H. and monotonicity of $\kindop{\laf}$} \\
        \eeq &
        \lapp{ \lfp \laop } \sqcap \laf
        \tag{by definition of $\kindop{\laf}$} \\
        \eeq &
        (\lfp \laop) \sqcap \laf 
        \\
        \llaordrev &
        \lfp \laop~.
	\end{align*}
    \emph{The case $\od$ limit ordinal.}
	\begin{align*}
        & \kindopfold{\laf}{\od}(\laf) \\
        \eeq &
          \inf \left\{ \kindopfold{\laf}{\oc}(\laf) \mathrel{\big|} \oc < \od \right\}
        \tag{by definition of $\kindopfold{\laf}{\od}(\laf)$} \\
        \llaordrev &
        \inf \left\{ \lfp \laop \right\}
        \tag{by I.H.} \\
        \eeq & \lfp \laop~.
	\end{align*}
	This completes the proof. \qed

\subsection{Proof of \Cref{cor:k_induction}}
\label{proof:cor:k_induction}
	\begin{align*}
		&\Phi\left(\kindopfold{\laf}{\lak}(\laf) \right) \llaord \laf \\
		\text{implies}\quad& \Phi\left(\kindopfold{\laf}{\lak}(\laf)\right) \llaord\kindopfold{\laf}{\lak}(\laf)  \tag{by \Cref{thm:k_induction_to_induction}}\\
		\text{implies}\quad &  \lfp \Phi\llaord\kindopfold{\laf}{\lak}(\laf)
		\tag{by Park induction} \\
		\text{implies}\quad & \lfp \Phi \llaord \laf~.
		\tag*{(by \Cref{lem:kindop_mon_chain}\ref{lem:kindop_mon_chain_2}) \qquad \qed} 
	\end{align*}%
	%

\subsection{Proof of \Cref{thm:kind-complete-for-unique}}
\label{proof:thm:kind-complete-for-unique}
    By the Knaster-Tarski theorem, we have $\ladown{\ob}(\top) = \gfp \laop$
    for some ordinal~$\ob$.
    We next observe that for this ordinal $\ob$, we have
    $\kindopfold{\laf}{\ob}(\laf) \laord \lfp \laop$:
    \begin{align*}
    \kindopfold{\laf}{\ob}(\laf) 
    \llaord & \ladown{\ob}(\laf)
    \tag{by \Cref{lem:kindop_mon_chain}\ref{lem:kindop_mon_chain_4}} \\
    \llaord & \ladown{\ob}(\top)
    \tag{by monotonicity of $\laop$} \\
    \eeq & \gfp \laop \eeq \lfp \laop
    \tag{by Knaster-Tarski theorem and the assumption}
    \end{align*}
    It follows that $\laf$ is $(\beta+1)$-inductive, since
    \begin{align*}
        \lapp{\kindopfold{\laf}{\beta}}
    	\llaord & \Phi\left(\lfp \Phi\right)
    	\tag{apply the above property and monotonicity of $\Phi$} \\
    	\eeq &  \lfp \Phi
    	\tag{by the fixed point property} \\
    	\llaord & \laf~.
    	\tag*{(by the implication's premise)\qquad\qed}
    \end{align*}

\subsection{Proof of \Cref{thm:latticed_vs_classical}}
\label{proof:thm:latticed_vs_classical}

\paragraph{Premises.}
We first recap our assumptions and introduce some useful notation.
We assume the setting from \Cref{sec:latticed-vs-classical}, in particular
\[
    \laop(X) \eeq \init ~\cup\, \succs(X)~,
\]
where $\succs(X) = \{ t' \mid \exists t \in X\colon (t,t') \in \transrel \}$.
We denote by $\mu$ interpretations that assign to every variable a state in $\stateset$; $\mu \models \varphi$ denotes that $\mu$ is a model of formula $\varphi$.
Moreover, we treat
$\initf(s)$, $\transrelf(s,s')$, and $\propsetf(s)$ as relational symbols 
whose interpretation is $\init$, $\transrel$, and $\propset$, respectively.\footnote{That is, by fixing an interpretation, we abstract from how $\initf(s)$, $\transrelf(s,s')$, and $\propsetf(s)$ are axiomatized. Moreover, by interpreting every variable as a state in $\stateset$, we abstract from common encodings of the state space in which a single state is given by the evaluation of a set of state variables (cf.~\cite{interpolating_strong_induction,DBLP:conf/vmcai/BjornerG15,sofware_k_induction}).}
Formally:
%
\begin{enumerate}
	\item $\mu \models \initf(s)$ iff $\mu(s) \in \initf$,
	\item $\mu \models \transrelf(s,s')$ iff $(\mu(s), \mu(s')) \in \transrel$, and
	\item $\mu \models \propsetf(s)$ iff $\mu(s)\in\propset$.
    \item $\forall s \in \stateset ~\exists s'\in\stateset \colon \  (s,s') \in \transrel$.
\end{enumerate}
Recall the formula \ref{form:kind_init} depicted below.
We denote by $\varphi_k(s_1,\ldots,s_k)$ the LHS of the implication and by $\psi_k(s_1,\ldots,s_k)$ the RHS of the implication, respectively.
	\begin{align*}
			\underbrace{\initf(s_1) \wedge \transrelf(s_1,s_2) \wedge \ldots \wedge \transrelf(s_{k-1},s_k)}_{\ddefeq \varphi_k} \implies \underbrace{\propsetf(s_1) \wedge \ldots \wedge \propsetf(s_k)}_{\ddefeq \psi_k }~.
            \tag{\ref{form:kind_init}}
	\end{align*}
Moreover, recall the formula \ref{form:kind_step}; as shown below, we define a shortcut for the LHS of the implication excluding the last transition.
	\begin{align*}
        \underbrace{
                \propsetf(s_1) \wedge \transrelf(s_1,s_2)
                \wedge \ldots \wedge \propset(s_k)
        }_{\ddefeq \pi_{k}}
            \wedge \transrelf(s_k,s_{k+1})
        \implies \propsetf(s_{k+1})~.
        \tag{\ref{form:kind_step}}
	\end{align*}

Finally, we define an auxiliary transformer capturing all successors of a given set of states that satisfy the property $\propset$:
    \[ 
        \Lambda\colon 2^{\stateset} \to 2^{\stateset}, 
        \qquad
        F ~\mapsto~ \succs(F) \cap \propset
    \]

\paragraph{Claim.}
For every natural number $k \geq 1$,
\begin{align*}
	\lapp{\kindopfold{\propset}{k-1}(\propset)} ~{}\subseteq{}~ \propset 
	\qiff
    \textnormal{formulae \eqref{form:kind_init} and \eqref{form:kind_step} are valid}~. 
\end{align*}
\begin{proof}[of the claim aka \Cref{thm:latticed_vs_classical}]
Let $k \geq 1$.
We rely on several lemmata, which are presented further below.
With this in mind, consider the following:
\begin{align*}
  &
  \lapp{\kindopfold{\propset}{k-1}(\propset)} ~{}\subseteq{}~ \propset
  \\
  \tiff &
  \lapp{\kindopfold{\propset}{k-1}(\propset)} ~{}\subseteq{}~ \propset
  \qand \lappdown{k}{\emptyset} ~{}\subseteq{}~ \propset 
  \tag{by \Cref{lem:lvc:00}} \\
  \tiff &
  \lapp{\lappdown{k-1}{\emptyset} \cup \auxdown{k-1}{\propset}} ~{}\subseteq{}~ \propset
  \tag{by \Cref{lem:lvc:01} and monotinicty of $\laop$} \\
  \tiff &
  \init \cup
  \succs\left(\lappdown{k-1}{\emptyset} \cup \auxdown{k-1}{\propset}\right) 
  ~{}\subseteq{}~ \propset
  \tag{by definition of $\laop$} \\
  \tiff &
  \init \cup
  \succs\left(\lappdown{k-1}{\emptyset}\right) \cup \succs\left(\auxdown{k-1}{\propset}\right) 
  ~{}\subseteq{}~ \propset
  \tag{$\succs(.)$ distributes over $\cup$} \\
  \tiff &
  \lappdown{k}{\emptyset} \cup \succs\left(\auxdown{k-1}{\propset}\right) 
  ~{}\subseteq{}~ \propset
  \tag{by definition of $\laop$} \\
  \tiff &
  \lappdown{k}{\emptyset} ~{}\subseteq{}~ \propset
  \qand 
  \succs\left(\auxdown{k-1}{\propset}\right) ~{}\subseteq{}~ \propset
  \\
  \tiff &
  \varphi_k \implies \psi_k \text{ is valid}
  \qand \succs\left(\auxdown{k-1}{\propset}\right) ~{}\subseteq{}~ \propset 
  \tag{by \Cref{lem:lvc:02}} \\
  \tiff &
  \varphi_k \implies \psi_k \text{ is valid}
  \tag{by \Cref{lem:lvc:03}} \\
  & \qand \pi_k \wedge \transrelf(s_k,s_{k+1}) \implies \propsetf(s_{k+1}) \text{ is valid} 
  \\
  \tiff &
  \textnormal{formulae \eqref{form:kind_init} and \eqref{form:kind_step} are valid}~. 
  \tag{by definition of $\varphi_k$, $\psi_k$, and $\pi_k$}
\end{align*}
\qed
\end{proof}
\begin{lemma}
\label{lem:lvc:00}
For all $k \geq 1$, 
$\lapp{\kindopfold{\propset}{k-1}(\propset)} \subseteq \propset$ implies $\lappdown{k}{\emptyset} \subseteq \propset$.
\end{lemma}
\begin{proof}
By \Cref{cor:k_induction}, $\lapp{\kindopfold{\propset}{k-1}(\propset)} \subseteq \propset$ implies
$\lfp \laop \subseteq \propset$.
Hence, for all $n \in \Nats$, we have $\lappdown{n}{\emptyset} \subseteq \propset$; in particular, for $n = k$.
\qed
\end{proof}
\begin{lemma}
\label{lem:lvc:01}
For all $k \geq 1$, if $\lappdown{k}{\emptyset} \subseteq \propset$, then
\[ \kindopfold{\propset}{k-1}(\propset) \eeq \lappdown{k-1}{\emptyset} \cup \auxdown{k-1}{\propset}~. \]
\end{lemma}
\begin{proof}
By induction on $k$. For $k = 1$, we have
\begin{align*}
\kindopfold{\propset}{1-1}(\propset) 
\eeq \propset 
\eeq \lappdown{1-1}{\emptyset} \cup \auxdown{1-1}{\propset}~.
\end{align*}
For $k > 1$, consider the following:
\begin{align*}
& \kindopfold{\propset}{k-1}(\propset) \\
\eeq &
\kindop{\propset}\left(\kindopfold{\propset}{k-2}(\propset)\right) 
\\
\eeq &
\lapp{\kindopfold{\propset}{k-2}(\propset)} \cap \propset
\tag{by definition of $\kindop{\propset}$} \\
\eeq &
\lapp{\lappdown{k-2}{\emptyset} \cup \auxdown{k-2}{\propset}} \cap \propset
\tag{by I.H.} \\
\eeq &
\left( \init \cup \succs\left(\lappdown{k-2}{\emptyset} \cup \auxdown{k-2}{\propset}\right)\right) \cap \propset
\tag{by definition of $\laop$} \\
\eeq &
\left( \init \cup \succs\left(\lappdown{k-2}{\emptyset}\right) \cup \succs\left(\auxdown{k-2}{\propset}\right)\right) \cap \propset
\tag{$\succs$ distributes over $\cup$} \\
\eeq &
\left( \lappdown{k-1}{\emptyset} \cup \succs\left(\auxdown{k-2}{\propset}\right)\right) \cap \propset
\tag{by definition of $\laop$} \\
\eeq &
\left( \lappdown{k-1}{\emptyset} \cap \propset \right) 
\cup \auxdown{k-1}{\propset}
\tag{by definition of $\auxdown{k-1}{\propset}$} \\
\eeq & 
\lappdown{k-1}{\emptyset} \cup \auxdown{k-1}{\propset}~.
\tag{by the assumption $\lappdown{k-1}{\emptyset} \subseteq \propset$}
\end{align*}
\qed
\end{proof}

\begin{lemma}
	\label{lem:lvc:02}
	For all $k \geq 1$,
	$\varphi_k \implies \psi_k$ is valid iff $\lappdown{k}{\emptyset} \subseteq \propset$.
\end{lemma}
\begin{proof}
	By induction on $k$. \\ \\
	\noindent
	\emph{Base case $k=1$.} We have %
	\begin{align*}
		& \varphi_1 \implies \psi_1~\text{is valid} \\%
		\text{iff}\quad &\forall \mu\colon~\mu \models \varphi_1~\text{implies}~\mu\models\psi_1 \\
		\text{iff}\quad& \forall \mu\colon~\mu \models \varphi_1~\text{implies}~\mu(s_1) \in \propset 
		\tag{by \Cref{lem:psi_equals_p}} \\
		\text{iff} \quad& \forall \mu\colon~\mu(s_1) \in \init ~\text{implies}~\mu(s_1) \in \propset \\
		\text{iff} \quad& \init \subseteq \propset \\
		\text{iff} \quad& \lappdown{1}{\emptyset} \subseteq \propset~. \\
	\end{align*}
	\emph{Induction step.}
	\begin{align*}
		& \varphi_{k+1} \implies \psi_{k+1}~\text{is valid} \\
		\text{iff}\quad& \varphi_{k+1} \implies \psi_{k} \wedge \propsetf(s_{k+1}) ~\text{is valid} \tag{by def.\ of $\psi_{k+1}$} \\
		\text{iff}\quad& (\varphi_{k+1} \implies \psi_{k}~\text{is valid})~\text{and}~(\varphi_{k+1} \implies \propsetf(s_{k+1})~\text{is valid}) \\
		\text{iff}\quad& (\varphi_{k} \wedge \transrelf(s_k,s_{k+1}) \implies \psi_{k}~\text{is valid})~\text{and}~(\varphi_{k+1} \implies \propsetf(s_{k+1})~\text{is valid})
		\tag{by def.\ of $\varphi_{{k+1}}$}\\
		\text{iff}\quad&(\transrelf(s_k,s_{k+1}) \implies (\varphi_{k} \implies \psi_{k})~\text{is valid})~\text{and}~(\varphi_{k+1} \implies \propsetf(s_{k+1})~\text{is valid})
		\\
		\text{iff}\quad& (\varphi_{k} \implies \psi_{k}~\text{is valid})~\text{and}~(\varphi_{k+1} \implies \propsetf(s_{k+1})~\text{is valid})
		\tag{since $\transrelf$ is total and $s_{k+1}$ does not occurr in $\varphi_k$ or $\psi_k$}\\
		\text{iff}\quad& \lappdown{k}{\emptyset} \subseteq \propset~\text{and}~(\varphi_{k+1} \implies \propsetf(s_{k+1})~\text{is valid})
		\tag{by I.H.}\\
		\text{iff}\quad& \lappdown{k}{\emptyset} \subseteq \propset~\text{and}~(\forall \mu \colon~\mu\models\varphi_{k+1} ~\text{implies}~ \mu\models\propsetf(s_{k+1}))
		\\
		\text{iff}\quad& \lappdown{k}{\emptyset} \subseteq \propset~ \\
		&\quad\text{and}~(\forall \mu \colon~\mu(s_{i+1}) \in \succsk{i}{\init}~\text{for all $i \in\{0,\ldots,k\}$}~\text{implies}~ \mu\models\propsetf(s_{k+1}))
		\tag{by Lemma~\ref{lem:phi_is_succs}}\\
		\text{iff}\quad& \lappdown{k}{\emptyset} \subseteq \propset~ \text{and}~(\forall \mu \colon~\mu(s_{k+1}) \in \succsk{k}{\init}~\text{implies}~ \mu\models\propsetf(s_{k+1}))\\
		\text{iff}\quad& \lappdown{k}{\emptyset} \subseteq \propset~ \text{and}~(\forall \mu \colon~\mu(s_{k+1}) \in \succsk{k}{\init}~\text{implies}~ \mu(s_{k+1})\in\propset )\\
		\text{iff}\quad& \lappdown{k}{\emptyset} \subseteq \propset~ \text{and}~(\succsk{k}{\init}\subseteq\propset )\\
		\text{iff}\quad& \lappdown{k+1}{\emptyset} \subseteq \propset~.
		\tag{by Lemma~\ref{lem:lapp_and_succ}}\\
		%
		%
		%
		%
		%
	\end{align*}

\end{proof}
%
%
%
%
%
\begin{lemma}
	\label{lem:lvc:03}
	For all $k \geq 1$,
	\[
	\pi_k \wedge \transrelf(s_k,s_{k+1}) \implies \propsetf(s_{k+1})
	\text{ is valid} 
	\qquad\text{iff}\qquad
	\succs(\auxdown{k-1}{\propset}) \,\subseteq\, \propset~.
	\]
\end{lemma}
\begin{proof}
	%
	%
	%
	We have
	\begin{align*}
		&\pi_{k} \wedge \transrelf(s_{k},s_{k+1}) \implies \propsetf(s_{k+1}) \text{ is valid} \\
		\text{iff}\quad & \forall \mu \colon~(\mu(s_{i+1}) \in \auxdown{i}{\propset}~\text{for all $i\in\{0,\ldots,k-1\}$}
		\tag{by \Cref{lem:auxdown_is_pi}} \\ &\quad \text{and}~\mu(s_{k+1}) \in \succs(\{\mu(s_{k})\}))~\text{implies}~\mu(s_{k+1}) \in \propset
		\\
		\text{iff}\quad & \forall \mu \colon~(\mu(s_{k}) \in \auxdown{k-1}{\propset} ~\text{and}~\mu(s_{k+1}) \in \succs(\{\mu(s_{k})\}))~\text{implies}~\mu(s_{k+1}) \in \propset
		\\
		\text{iff}\quad & \succs(\auxdown{k}{\propset}) \subseteq \propset~.
	\end{align*}
	
\end{proof}
\begin{lemma}
	\label{lem:psi_equals_p}
	For all $k \geq 1$, $\mu \models \psi_{k}$ iff $\mu(s_1), \ldots, \mu(s_k) \in \propset$.
\end{lemma}
\begin{proof}
	Immediate by assumption (3).
\end{proof}
\begin{lemma}
	\label{lem:phi_is_succs}
	For all $k \geq 1$, $\mu \models \varphi_{k}$ iff $\mu(s_{i+1}) \in \succsk{i}{\init}$ 
	\mbox{for all $i \in \{0,\ldots,k-1\}$.}
\end{lemma}
\begin{proof}
	By induction on $k$. \\ \\
	\noindent
	\emph{Base case $k=1$.} We have
	\begin{align*}
		& \mu \models \varphi_1 \\
		\text{iff} \quad & \mu \models \initf(s_1) 
		\tag{by def.\ of $\varphi_1$} \\
		\text{iff} \quad & \mu(s_1) \in \init \\
		\text{iff} \quad & \mu(s_1) \in \succsk{0}{\init}~. 
	\end{align*}
	\emph{Induction step.} We have
	\begin{align*}
		& \mu\models \varphi_{k+1} \\
		\text{iff}\quad& \mu \models \varphi_k \wedge \transrelf(s_k,s_{k+1}) \\
		\text{iff} \quad & \mu \models \varphi_k ~\text{and}~\mu\models\transrelf(s_k,s_{k+1}) \\
		\text{iff} \quad & (\mu(s_{i+1}) \in \succsk{i}{\init}~\text{for all $i \in \{0,\ldots,k-1\}$})
		\tag{by I.H.}\\
		&\quad \text{and}~\mu\models\transrelf(s_k,s_{k+1})
		\\
		\text{iff} \quad & (\mu(s_{i+1}) \in \succsk{i}{\init}~\text{for all $i \in \{0,\ldots,k-1\}$})\\
		&\quad \text{and}~\mu(s_{k+1}) \in \succs(\{\mu(s_k)\}) \\
		\text{iff} \quad & \mu(s_{i+1}) \in \succsk{i}{\init}~\text{for all $i \in \{0,\ldots,k\}$}~.
	\end{align*}
\end{proof}
\begin{lemma}
	\label{lem:lapp_and_succ}
	For all $k \geq 1$, $\lappdown{k}{\emptyset} = \lappdown{k-1}{\emptyset} \cup \succsk{k-1}{\init}$.
\end{lemma}
\begin{proof}
	By induction on $k$.
\end{proof}
\begin{lemma}
	\label{lem:auxdown_is_pi}
For all $k \geq 1$, $\mu \models\pi_k$ iff $\mu(s_{i+1}) \in \auxdown{i}{\propset}$ for all \mbox{$i \in\{0,\ldots,k-1\}$~.}
\end{lemma}
\begin{proof}
   By induction on $k$. \\ \\
   \noindent
   \emph{Base case $k=1$.} We have
   \begin{align*}
   	  &\mu \models \pi_1 \\
   	  \text{iff}\quad &\mu\models\propsetf(s_1)\\
   	  \text{iff}\quad & \mu(s_1) \in\propset \\
   	   \text{iff}\quad & \mu(s_1) \in\auxdown{0}{\propset}~.
   \end{align*}
%
  \emph{Induction step.} We have
     \begin{align*}
     	&\mu \models \pi_{k+1} \\
     	\text{iff}\quad & \mu \models \pi_k \wedge \transrelf(s_k,s_{k+1}) \wedge \propsetf(s_{k+1}) \\
     	\text{iff}\quad &  \mu(s_{i+1}) \in \auxdown{i}{\propset}~\text{for all}~i \in\{0,\ldots,k-1\}~\text{and}~\mu\models \transrelf(s_k,s_{k+1}) \wedge \propsetf(s_{k+1})
     	\tag{by I.H.} \\
     	\text{iff}\quad&\mu(s_{i+1}) \in \auxdown{i}{\propset}~\text{for all}~i \in\{0,\ldots,k-1\}~\text{and}~\mu(s_{k+1}) \in \succs(\{\mu(s_k)\})\cap\propset \\
     	\text{iff}\quad&\mu(s_{i+1}) \in \auxdown{i}{\propset}~\text{for all}~i \in\{0,\ldots,k-1\}~\text{and}~\mu(s_{k+1}) \in \succs(\auxdown{k-1}{\propset}) \cap \propset \\
     	\text{iff}\quad&\mu(s_{i+1}) \in \auxdown{i}{\propset}~\text{for all}~i \in\{0,\ldots,k-1\}~\text{and}~\mu(s_{k+1}) \in \auxdown{k}{\propset}
     	\tag{since for $k\geq 1$, $\auxdown{k}{\propset} = \auxdown{1}{\auxdown{k-1}{\propset}} = \succs(\auxdown{k-1}{\propset}) \cap \propset$} \\
     	\text{iff}\quad&\mu(s_{i+1}) \in \auxdown{i}{\propset}~\text{for all}~i \in\{0,\ldots,k\} ~.
     \end{align*}
\end{proof}
%
%
%
%
%
%
%
%
%
%
%
%
%
%
%
%
%
%
%

\subsection{Formal Definition of Linear Rescaling}
\label{app:linear-rescaling}

Let $\fh \in \Elin$ be a linear expectation.
Moreover, let $c \in \PosRats$ be some rational constant.
For $c = 0$, we define the rescaling as $0 \cdot \fg \defeq 0$.
For $c > 0$, the rescaling $c \cdot \fh$ is given by the table below.

\begin{center}
\begin{tabular}{rr}
\toprule
  $\fh$ & $c \cdot \fh$ \\
  \midrule
  $\infty$ & $\infty$ \\
  $e~(\neq \infty)$ & $c \cdot e$ \\
  $\iverson{\bb} \cdot \fhp$ & $\iverson{\bb} \cdot (c \cdot \fhp)$ \\
  $\fhp + \fhp'$ & ~~~~~~$c \cdot \fhp + c \cdot \fhp'$\\
\bottomrule
\end{tabular}
\end{center}

\subsection{Proof of \Cref{lem:gnf}}
\label{app:gnf}

\paragraph{Claim.}
Every linear expectation $\fh \in \Elin$ can effectively be transformed 
into an equivalent linear expectation $\gnf{\fh} \equiv \fh$ in guarded normal form.
\begin{proof}
  Let $\fh \in \Elin$.
  As shown in \cite[Lemma A.2]{relatively_complete_prob_progs}, $\fh$ 
  can effectively be transformed into an equivalent (linear) expectation 
  $\fhp$ of the form 
  \begin{align*}
    \fhp \eeq \sum\nolimits_{i = 1}^{m} \iverson{\bp_i} \cdot \eainfty_i~,
  \end{align*}
  that is not necessarily in GNF (as the guards $\bp_1,\ldots,\bp_n$ do not necessarily partitition the state space).
    We can then construct an equivalent linear expectation $\gnf{\fh}$ in guarded normal form as follows:
  \begin{align*}
	\gnf{\fh} \ddefeq 
	\sum_{\left( (\br_1, \eoinfty_1),\ldots,(\br_m, \eoinfty_m)\right)\, \in\, \bigtimes_{i=1}^{m} \left\{ (\bp_i,\eainfty_i), (\neg\bp_i, 0) \right\}}
      \,
      \Big[ \underbrace{\bigwedge\nolimits_{i=1}^m \br_i}_{\eeq \bb_i} \Big]
      \,\cdot\,
      \underbrace{ \Big( \sum\nolimits_{i=1}^m \eoinfty_i \Big)}_{\eeq \eeinfty_i}~,
  \end{align*}
  where we define $\infty + \eainfty = \eainfty + \infty = \infty$. 
\qed
\end{proof}%

\subsection{Proof of $\smtentails{\fh}{\fh'}$}
\label{app:entail-proof}
We show that
$\fh \preceq \fhp$ holds iff $\smtentails{\fh}{\fhp}$ is \emph{un}satisfiable:
\begin{align*}
	&\fh \ppreceq \fhp \\
	\text{iff} \qquad & \gnf{\fh} \ppreceq \gnf{\fhp} 
	\tag{by \Cref{lem:gnf}} \\
	\text{iff} \qquad & 
	\sum\limits_{i=1}^{n} \iverson{\bb_i}\cdot \eeinfty_i 
	\ppreceq
	\sum\limits_{i=1}^{m} \iverson{\bp_i}\cdot \eainfty_i
	\tag{by definiton of $\gnf{.}$} \\
	\text{iff} \qquad & 
	\forall \pstate \in \States\colon
	\sum\limits_{i=1}^{n} (\iverson{\bb_i}\cdot \eeinfty_i)(\pstate)
	\lleq
	\sum\limits_{i=1}^{m} (\iverson{\bp_i}\cdot \eainfty_i)(\pstate)
	\tag{by definiton of $\preceq$} \\
	\text{iff} \qquad & 
	\forall \pstate \in \States\colon \pstate \models
	\bigwedge_{i=1}^n \left( \neg \bb_i \vee \eeinfty_i \leq \sum\limits_{i=1}^{m} \iverson{\bp_i}(\pstate)\cdot \eainfty_i \right) 
	\tag{$\pstate \models \bb_i$ for exactly one $i$} \\
	\text{iff} \qquad & 
	\forall \pstate \in \States\colon \pstate \models
	\bigwedge_{i=1}^n \bigwedge_{j=1}^{m} \neg \bb_i \vee \neg \bp_j \vee \eeinfty_i \leq \eainfty_i
	\tag{$\pstate \models \bp_j$ for exactly one $j$} \\
	\text{iff} \qquad & 
	\neg \exists \pstate \in \States\colon \pstate \models
	\bigvee_{i=1}^n \bigvee_{\substack{j=1}}^{m} \bb_i \wedge \bp_j \wedge \eeinfty_i > \eainfty_i
	\tag{by de Morgan's law} \\
	\text{iff} \qquad & 
	\neg \exists \pstate \in \States\colon \pstate \models
	\bigvee_{i=1}^n \bigvee_{\substack{j=1, \\ \eainfty_j \neq \infty}}^{m} \bb_i \wedge \bp_j \wedge \tosmt{\eeinfty_i} > \eainfty_i
	\tag{encoding of $\infty$} \\
	\text{iff} \qquad & \smtentails{\fh}{\fhp}~\text{is unsatisfiable}~.
	\tag{by definition of $\smtentails{\fh}{\fhp}$}
\end{align*}

\subsection{$(\omega+1)$-Inductivity of $2x+1$}
We show for every $n\geq 1$ that
\begin{align*}
   \kindopfold{2x+1}{n}(2x+1) \eeq & \iverson{c\neq 1}\cdot x + \iverson{c=1} \cdot \iverson{x=0} \\
   &\quad + \iverson{c=1} \cdot \iverson{x \geq 1} \cdot \Big(
   \big(  2 - \sum_{i=1}^{n-1} \frac{1}{2^i} \big) \cdot x + 1 +  \frac{n-1}{2^{n-1}} \Big)~.
\end{align*}
From that, we get
\[
    \kindopfold{2x+1}{\omega}(2x+1) \eeq \sup_n \kindopfold{2x+1}{n}(2x+1)  \eeq \iverson{c\neq 1}\cdot x +  \iverson{c=1} \cdot (x+1) 
\]
and thus
\[
  \laop(\kindopfold{2x+1}{\omega}(2x+1)) \lleq 2x+1~,
\]
rendering $2x+1$ an $(\omega+1)$-inductive invariant. 

Furthermore, from \Cref{lem:omega_ind_phi} and \Cref{thm:closed_form_psi} it follows for every $n\geq1$ that $\laop(\kindopfold{2x+1}{n}(2x+1)) \not\preceq 2x+1$ since $\laop(\kindopfold{2x+1}{n}(2x+1))(\pstate) >1$ for every $\pstate \in \States$ with $\pstate(x)=0$, i.e., there is no $n\geq 1$ such that $2x+1$ is $n+1$-inductive (and thus also not $1$-inductive). 
\begin{lemma}
	\label{lem:omega_ind_phi}
	For $n \in\Nats$, if 
	\begin{align*}
	    \kindopfold{2x+1}{n}(2x+1) \eeq & \iverson{c\neq 1}\cdot x + \iverson{c=1} \cdot \iverson{x=0} \\
	    &\quad + \iverson{c=1} \cdot \iverson{x \geq 1} \cdot \Big(
	    \big(  2 - \sum_{i=1}^{n-1} \frac{1}{2^i} \big) \cdot x + 1 +  \frac{n-1}{2^{n-1}} \Big) ~
	\end{align*}
	then
	\begin{align*}
	   \laop(\kindopfold{2x+1}{n}(2x+1) ) \eeq &\iverson{c\neq 1}\cdot x \\
	   &\quad + \iverson{c=1} \cdot \Big(
	   \big(  2 - \sum_{i=1}^{n} \frac{1}{2^i} \big) \cdot x + 1 +  \frac{n}{2^{n}} \Big) ~
	\end{align*}
\end{lemma}
\begin{proof}
	We have 
	\begin{align*}
		&\laop(\kindopfold{2x+1}{n}(2x+1) ) \\
		\eeq & \iverson{c\neq1} \cdot x \\
		&\quad \iverson{c=1} \cdot \frac{1}{2}\cdot (\kindopfold{2x+1}{n}(2x+1)\subst{c}{0} + \kindopfold{2x+1}{n}(2x+1)\subst{x}{x+1} ) \\
		\eeq & \iverson{c\neq1} \cdot x \\
		&\quad + \iverson{c=1}\cdot \frac{1}{2}\cdot\big( 
		    x + (  2 - \sum_{i=1}^{n} \frac{1}{2^i} ) \cdot (x+1) + 1 +  \frac{n-1}{2^{n-1}}  \big) \\
		 \eeq & \iverson{c\neq1} \cdot x \\
		 &\quad + \iverson{c=1}\cdot \big( 
		 \frac{1}{2}\cdot x + (  1 - \sum_{i=1}^{n} \frac{1}{2^{i+1}} ) \cdot (x+1) + \nicefrac{1}{2} +  \frac{n-1}{2^{n}}  \big)   \\
		 \eeq & \iverson{c\neq1} \cdot x \\
		 &\quad + \iverson{c=1}\cdot \big( 
		 \frac{1}{2}\cdot x + (  1 - \sum_{i=1}^{n} \frac{1}{2^{i+1}} ) \cdot x + (  1 - \sum_{i=1}^{n} \frac{1}{2^{i+1}} ) + \frac{1}{2} +  \frac{n-1}{2^{n}}  \big)   \\
		  \eeq & \iverson{c\neq1} \cdot x \\
		  &\quad + \iverson{c=1}\cdot \big( 
		    (\underbrace{\frac{1}{2} + 1 - \sum_{i=1}^{n} \frac{1}{2^{i+1}} ) \cdot x}_{=(  2 - \sum_{i=1}^{n} \frac{1}{2^i})\cdot x} + \underbrace{(  1 - \sum_{i=1}^{n} \frac{1}{2^{i+1}} ) + \frac{1}{2} +  \frac{n-1}{2^{n}}}_{= 1+\frac{n}{2^n}} \big)   \\
		 \eeq & \iverson{c\neq 1}\cdot x \\
		 &\quad + \iverson{c=1} \cdot \Big(
		 \big(  2 - \sum_{i=1}^{n} \frac{1}{2^i} \big) \cdot x + 1 +  \frac{n}{2^{n}} \Big)~.
   \end{align*}
\end{proof}
%
%
%
%
%
%
\begin{theorem}
	\label{thm:closed_form_psi}
	For every $n\geq1$, 
	\begin{align*}
	\kindopfold{2x+1}{n}(2x+1) \eeq & \iverson{c\neq 1}\cdot x + \iverson{c=1} \cdot \iverson{x=0} \\
	&\quad + \iverson{c=1} \cdot \iverson{x \geq 1} \cdot \Big(
	\big(  2 - \sum_{i=1}^{n-1} \frac{1}{2^i} \big) \cdot x + 1 +  \frac{n-1}{2^{n-1}} \Big)~.
	\end{align*}
\end{theorem}
\begin{proof}
	By induction on $n$. \\ \\
	\noindent
	\emph{Base case $n=1$.} We have
	\begin{align*}
	   &\kindopfold{2x+1}{n}(2x+1)  \\
	   \eeq & \laop(2x+1) \emin 2x+1 \\
	   \eeq& \iverson{c\neq 1} \cdot x \\
	       & \quad + \iverson{c=1}\cdot \frac{1}{2}\cdot ( (2x+1)\subst{c}{0} + (2x+1)\subst{x}{x+1}))\emin 2x+1 \\
	    \eeq& \iverson{c\neq 1} \cdot x + \iverson{c=1}\cdot \frac{1}{2}\cdot ( 2x+1 + 2x+3))\emin 2x+1 \\
	   \eeq& \iverson{c\neq 1} \cdot x + \iverson{c=1}\cdot \frac{1}{2}\cdot ( 4x+4))\emin 2x+1 \\
	   \eeq& \iverson{c\neq 1} \cdot x  + \iverson{c=1}\cdot  ( 2x+2))\emin 2x+1 \\
	   \eeq& \iverson{c\neq 1} \cdot x  + \iverson{c=1}\cdot  ( 2x+1))\emin 2x+1 \\
	   \eeq& \iverson{c\neq 1}\cdot x + \iverson{c=1} \cdot \iverson{x=0} \\
	   &\quad + \iverson{c=1} \cdot \iverson{x \geq 1} \cdot \Big(
	   \big(  2 - \sum_{i=1}^{0} \frac{1}{2^i} \big) \cdot x + 1 +  \frac{0}{2^{n-1}} \Big)~.
	\end{align*}
   \emph{Induction step.} We have
   \begin{align*}
   	   &\kindopfold{2x+1}{n+1}(2x+1)  \\
   	   \eeq & \kindop{2x+1}(\kindopfold{2x+1}{n}(2x+1) ) \\
   	   \eeq & \laop(\kindopfold{2x+1}{n}(2x+1)) \emin 2x+1 \\
   	   \eeq & \iverson{c\neq 1}\cdot x + \iverson{c=1}\cdot\iverson{x=0}
   	      	   \tag{by \Cref{lem:omega_ind_phi}} \\
   	   &\quad + \iverson{c=1}\cdot\iverson{x>0} \cdot \Big(
   	   \big(  2 - \sum_{i=1}^{n} \frac{1}{2^i} \big) \cdot x + 1 +  \frac{n}{2^{n}} \Big)~.
   	\end{align*}
\end{proof}

\section{Benchmarks}
\label{app:benchmarks}


\subsection{Brp}

A $\pgcl$ variant of the bounded retransmission protocol~\cite{DBLP:conf/papm/DArgenioJJL01,DBLP:conf/types/HelminkSV93}. The goal is to transmit $\vartosend$ number of packages via an unreliable channel allowing for at most $\varmaxfailed$ number of retransmissions per package. Variable $\vartotalfailed$ keeps track of the total number of failed attempts to send a package. We verified upper bounds on the expected outcome of $\vartotalfailed$ (variants $1$-$4$). In doing so, we bound the number of packages to send by $4$ (variant $1$) until $70$ (variant $4$) while keeping $\varmaxfailed$ \emph{unbounded}, i.e., we still verify an infinite-state system. 
We notice that $k > 1$ is required for proving any of the candidate bounds; 
for up to $k = 11$, $\tool$ manages to prove non-trivial bounds within a few seconds.
However, unsurprisingly, the complexity increases rapidly with larger~$k$.
While $\tool$ can prove variant 4, it needs to increase $k$ to 23; we observe that, unsurprisingly, the complexity grows rapidly both in terms of formulae and in terms of runtime with increased $k$.
Furthermore, variants $5$ -- $7$ correspond to (increasing) incorrect candidate bounds that are refuted (or time out)
when not imposing any restriction on $\vartosend$.

\begin{lstlisting}
	# The number of total packages to send
	nat toSend;
	
	# Number of packages sent
	nat sent;
	
	# The maximal number of retransmission tries
	nat maxFailed;
	
	# The number of failed retransmission tries
	nat failed;
	
	nat totalFailed;
	
	while(failed < maxFailed & sent < toSend){
		{
			# Transmission of current packages successful
			failed := 0;
			sent := sent + 1;
		}
		[0.9]
		{
			# Transmission not successful
			failed := failed +1;
			totalFailed := totalFailed + 1;
		}
	}
\end{lstlisting}

Preexpectations for the different variants:
\begin{itemize}
	\item $\iverson{toSend \leq 4}\cdot (totalFailed + 1) + \iverson{toSend > 4}\cdot \infty$
	\item $\iverson{toSend \leq 10}\cdot (totalFailed + 3) + \iverson{toSend > 10}\cdot \infty$
	\item $\iverson{toSend \leq 20}\cdot (totalFailed + 3) + \iverson{toSend > 20}\cdot \infty$
	\item $\iverson{toSend \leq 70}\cdot (totalFailed + 20) + \iverson{toSend > 70}\cdot \infty$
	\item $totalFailed +1$
	\item $totalFailed + 1.5$
	\item $totalFailed + 3$
\end{itemize}

\subsection{Geo}

The geometric loop $\cgeo$ from \Cref{ex:pgcl_kind_bmc}. 
We verify that $c+1$ upper bounds the expected value of $c$ for every initial state (variant $1$). Furthermore, we refute the candidates $c+0.99$ and $c+0.999999999999$ (variants $2$--$3$).

\begin{lstlisting}
	nat c;
	nat f;
	
	while(f=1){
		{f := 0}[0.5]{c := c+1}
	}
\end{lstlisting}

Preexpectations for the different variants:
\begin{itemize}
	\item $c+1$
	\item $c+0.99$
	\item $c+0.999999999999$
\end{itemize}

\subsection{Rabin}

A $\pgcl$ variant of Rabin's mutual exclusion algorithm~\cite{DBLP:conf/podc/KushilevitzR92} taken from~\cite{DBLP:journals/tcs/HurdMM05}. We verify $\nicefrac{2}{3}$ as an upper bound on the probability of obtaining a unique winning process when bounding the number of participating processes by $2$ (variant~$1$) up to $4$ (variant $3$). Furthermore, we refute $\nicefrac{1}{3}$ (variant $4$) and $\nicefrac{3}{5}$  (variant $5$) without restricting the number of participating processes. 

\begin{lstlisting}
nat i;
nat n; 
nat d;

nat phase; # Initially 0

while(1<i || phase=1){
	if(phase=0){
		n:=i;
		phase:=1;
	}{
		if(0<n){
			{d:=0}[0.5]{d:=1};
			i:=i-d;
			n:=n-1;
		}{ #leave inner loop
			phase:=0;
		}
	}
}
\end{lstlisting}

Preexpectations for the different variants:
\begin{itemize}
	\item $\iverson{1<i \wedge i< 2 \wedge phase=0}\cdot \frac{2}{3} + \iverson{\neg (1<i \wedge i< 2 \wedge phase=0)}\cdot 1$
	\item $\iverson{1<i \wedge i< 3 \wedge phase=0}\cdot \frac{2}{3} + \iverson{\neg (1<i \wedge i< 3 \wedge phase=0)}\cdot 1$
	\item  $\iverson{1<i \wedge i< 4 \wedge phase=0}\cdot \frac{2}{3} + \iverson{\neg (1<i \wedge i< 4 \wedge phase=0)}\cdot 1$
	\item $\iverson{1<i \wedge phase=0}\cdot \frac{1}{3} + \iverson{\neg (1<i \wedge phase=0)}\cdot 1$
	\item $\iverson{1<i \wedge phase=0}\cdot 0.6 + \iverson{\neg (1<i \wedge phase=0)}\cdot 1$
\end{itemize}

\subsection{Unif\_gen}

A $\pgcl$ variant of the algorithm from~\cite{DBLP:journals/corr/abs-1304-1916} for generating a discrete uniform distribution over some interval $\{ \varfont{elow},\varfont{elow}+1, \ldots,\varfont{ehigh} \}$ using fair coin flips only. We verify that $\nicefrac{1}{n}$ is an upper bound on the probability of sampling a particular element from \emph{any} such interval when bounding the number of elements by $n=2$ (variant $1$) up to $n=6$ (variant $5$) .

\begin{lstlisting}
	
	nat elow;
	nat ehigh; # Initially elow <= ehigh
	nat n; # Initially ehigh-elow + 1
	nat v; # Initially 1
	nat c; # Initially 0; the result
	nat running; # Initially 0
	
	nat i; # auxiliary variable for array positions in specifications
	
	while(running = 0){
		
		v := 2*v;
		{c := 2*c+1}[0.5]{c := 2*c};
		if((not (v<n))){
			if((not (n=c)) & (not (n<c))){ # terminate
				running := 1
			}{
				v := v-n;
				c := c-n;
			}
		}{
			skip
		}
		
		# On termination, determine correct index
		if((not (running = 0))){
			c := elow + c;
		}{
			skip
		}
	}
	
\end{lstlisting}

Preexpectations for the different variants:
\begin{itemize}
	\item $\iverson{elow+1=ehigh \wedge n=ehigh-elow+1 \wedge v=1 \wedge c=0 \wedge elow \leq i \leq ehigh} \cdot 0.5 + \iverson{\neg (elow+1=ehigh \wedge n=ehigh-elow+1 \wedge v=1 \wedge c=0 \wedge elow \leq i \leq ehigh)}\cdot 1$
	\item $\iverson{elow+2=ehigh \wedge n=ehigh-elow+1 \wedge v=1 \wedge c=0 \wedge elow \leq i \leq ehigh} \cdot \nicefrac{1}{3} + \iverson{\neg (elow+2=ehigh \wedge n=ehigh-elow+1 \wedge v=1 \wedge c=0 \wedge elow \leq i \leq ehigh)}\cdot 1$
	\item $\iverson{elow+3=ehigh \wedge n=ehigh-elow+1 \wedge v=1 \wedge c=0 \wedge elow \leq i \leq ehigh} \cdot \nicefrac{1}{4} + \iverson{\neg (elow+3=ehigh \wedge n=ehigh-elow+1 \wedge v=1 \wedge c=0 \wedge elow \leq i \leq ehigh)}\cdot 1$
	\item $\iverson{elow+4=ehigh \wedge n=ehigh-elow+1 \wedge v=1 \wedge c=0 \wedge elow \leq i \leq ehigh} \cdot \nicefrac{1}{5} + \iverson{\neg (elow+4=ehigh \wedge n=ehigh-elow+1 \wedge v=1 \wedge c=0 \wedge elow \leq i \leq ehigh)}\cdot 1$
	\item $\iverson{elow+5=ehigh \wedge n=ehigh-elow+1 \wedge v=1 \wedge c=0 \wedge elow \leq i \leq ehigh} \cdot \nicefrac{1}{6} + \iverson{\neg (elow+5=ehigh \wedge n=ehigh-elow+1 \wedge v=1 \wedge c=0 \wedge elow \leq i \leq ehigh)}\cdot 1$
\end{itemize}

\subsection{2drwalk}

\begin{lstlisting}
	nat x;
	nat y;
	nat d;
	nat n;
	while (d < n) {
		if (0 < x) {
			if (0 < y) {
				{
					x := x + 2;
					d := d + 2;
				} [1/4] {
					{
						y := y + 2;
						d := d + 2;
					} [1/3] {
						{
							x := x - 1;
							d := d - 1;
						} [1/2] {
							y := y - 1;
							d := d - 1;
						}
					}
				}
			} else {
				if (y < 0) {
					{
						x := x + 2;
						d := d + 2;
					} [1/4] {
						{
							y := y + 1;
							d := d - 1;
						} [1/3] {
							{
								x := x - 1;
								d := d - 1;
							} [1/2] {
								y := y - 2;
								d := d + 2;
							}
						}
					}
				} else {
					{
						x := x + 2;
						d := d + 2;
					} [1/4] {
						{
							y := y + 1;
							d := d + 1;
						} [1/3] {
							{
								x := x - 1;
								d := d - 1;
							} [1/2] {
								y := y - 1;
								d := d + 1;
							}
						}
					}
				}
			}
		} else {
			if (x < 0) {
				if (0 < y) {
					{
						x := x + 1;
						d := d - 1;
					} [1/4] {
						{
							y := y + 2;
							d := d + 2;
						} [1/3] {
							{
								x := x - 2;
								d := d + 2;
							} [1/2] {
								y := y - 1;
								d := d - 1;
							}
						}
					}
				} else {
					if (y < 0) {
						{
							x := x + 1;
							d := d - 1;
						} [1/4] {
							{
								y := y + 1;
								d := d - 1;
							} [1/3] {
								{
									x := x - 2;
									d := d + 2;
								} [1/2] {
									y := y - 2;
									d := d + 2;
								}
							}
						}
					} else {
						{
							x := x + 1;
							d := d - 1;
						} [1/4] {
							{
								y := y + 1;
								d := d + 1;
							} [1/3] {
								{
									x := x - 2;
									d := d + 2;
								} [1/2] {
									y := y - 1;
									d := d + 1;
								}
							}
						}
					}
				}
			} else {
				if (0 < y) {
					{
						x := x + 1;
						d := d + 1;
					} [1/4] {
						{
							y := y + 2;
							d := d + 2;
						} [1/3] {
							{
								x := x - 1;
								d := d + 1;
							} [1/2] {
								y := y - 1;
								d := d - 1;
							}
						}
					}
				} else {
					if (y < 0) {
						{
							x := x + 1;
							d := d + 1;
						} [1/4] {
							{
								y := y + 1;
								d := d - 1;
							} [1/3] {
								{
									x := x - 1;
									d := d + 1;
								} [1/2] {
									y := y - 2;
									d := d + 2;
								}
							}
						}
					} else {
						{
							x := x + 1;
							d := d + 1;
						} [1/4] {
							{
								y := y + 1;
								d := d + 1;
							} [1/3] {
								{
									x := x - 1;
									d := d + 1;
								} [1/2] {
									y := y - 1;
									d := d + 1;
								}
							}
						}
					}
				}
			}
		}
		tick(1);
	}
\end{lstlisting}

\subsection{bayesian}

\begin{lstlisting}
	
	nat i;
	nat d;
	nat s;
	nat l;
	nat g;
	nat n;
	while (0 < n) {
		i := 1 : 3/10 + 0 : 7/10;
		tick(1);
		d := 1 : 2/5 + 0 : 3/5;
		tick(1);
		if ((i < 1 & d < 1)) {
			g := 1 : 7/10 + 0 : 3/10;
			tick(1);
		} else {
			if ((i < 1 & 0 < d)) {
				g := 1 : 19/20 + 0 : 1/20;
				tick(1);
			} else {
				if ((0 < i & d < 1)) {
					g := 1 : 1/10 + 0 : 9/10;
					tick(1);
				} else {
					g := 1 : 1/2 + 0 : 1/2;
					tick(1);
				}
			}
		}
		if (i < 1) {
			s := 1 : 1/20 + 0 : 19/20;
			tick(1);
		} else {
			s := 1 : 4/5 + 0 : 1/5;
			tick(1);
		}
		if (g < 1) {
			l := 1 : 1/10 + 0 : 9/10;
			tick(1);
		} else {
			l := 1 : 3/5 + 0 : 2/5;
			tick(1);
		}
		n := n - 1;
	}
	
	
\end{lstlisting}

\subsection{ber}

\begin{lstlisting}
	
	nat x;
	nat n;
	nat r;
	while (x < n) {
		r := 1 : 1/2 + 0 : 1/2;
		x := x + r;
		tick(1);
	}
	
\end{lstlisting}

\subsection{C4B\_t303}

\begin{lstlisting}

nat x;
nat y;
nat t;
nat r;
while (0 < x) {
	r := 1 : 1/3 + 2 : 1/3 + 3 : 1/3;
	x := x - r;
	t := x;
	x := y;
	y := t;
	tick(1);
}

\end{lstlisting}

\subsection{condand}

\begin{lstlisting}
	
	nat n;
	nat m;
	while ((0 < n & 0 < m)) {
		{n := n - 1;} [1/2] {m := m - 1;}
		tick(1);
	}
	
\end{lstlisting}

\subsection{fcall}

\begin{lstlisting}
	
nat x;
nat n;
nat r;
while (x < n) {
	r := 0 : 1/2 + 1 : 1/2;
	x := x + r;
	tick(1);
}

	
\end{lstlisting}

\subsection{hyper}

\begin{lstlisting}
	
	nat x;
	nat n;
	nat r;
	while ((x + 2 <= n)) {
		r := 0 : 351/435 + 1 : 81/435 + 2 : 3/435;
		x := x + r;
		tick(1);
	}
	
	
\end{lstlisting}

\subsection{linear01}

\begin{lstlisting}
	
nat x;
while (2 <= x) {
	{x := x - 1;} [1/3] {x := x - 2;}
	tick(1);
}
	
\end{lstlisting}

\subsection{prdwalk}

\begin{lstlisting}
	
nat x;
nat n;
nat r;
while (x < n) {
	{
		r := 0 : 1/3 + 1 : 1/3 + 2 : 1/3;
		x := x + r;
	} [1/2] {
		r := 0 : 1/6 + 1 : 1/6 + 2 : 1/6 + 3 : 1/6 + 4 : 1/6 + 5 : 1/6;
		x := x + r;
	}
	tick(2);
}
	
\end{lstlisting}

\subsection{prspeed}

\begin{lstlisting}
	
nat x;
nat y;
nat m;
nat n;
while ((x + 3 <= n)) {
	if (y < m) {
		{ y := y + 1; } [1/2] {y := y + 0;}
	} else {
		{ x := x + 0; } [1/4] {
			{ x := x + 1; } [1/3] {
				{ x := x + 2; } [1/2] {
					x := x + 3;
				}
			}
		}
	}
	tick(1);
}
	
\end{lstlisting}

\subsection{race}

\begin{lstlisting}
	
nat h;
nat t;
nat r;
nat ticks;
while (h <= t) {
	t := t + 1;
	{
		r := 0 : 1/11 + 1 : 1/11 + 2 : 1/11 + 3 : 1/11 + 4 : 1/11 + 5 : 1/11 + 6 : 1/11 + 7 : 1/11 + 8 : 1/11 + 9 : 1/11 + 10 : 1/11;
		h := h + r;
	} [1/2] { h := h + 0; }
	tick(1);
}
	
\end{lstlisting}

\subsection{rdwalk}

\begin{lstlisting}
	
nat x;
nat n;
while (x < n) {
	{x := x + 2;} [1/2] {x := x - 1;}
	tick(1);
}
	
\end{lstlisting}

\subsection{sprdwalk}

\begin{lstlisting}
	
nat x;
nat n;
nat r;
while (x < n) {
	r := 0 : 1/2 + 1 : 1/2;
	x := x + r;
	tick(1);
}

	
\end{lstlisting}

%
%


\end{document}